\documentclass[aps,pre,preprint,groupedaddress,showpacs]{revtex4-1}
\usepackage{bm}
\usepackage{amsmath,amssymb}
\usepackage{graphicx}

\begin{document}


\title{Global Dynamics of a Stochastic Neuronal Oscillator}


\author{Takanobu Yamanobe}
\email[]{yamanobe@med.hokudai.ac.jp}
\affiliation{Hokkaido University School of Medicine, North 15, West 7, Kita-ku, Sapporo 060-8638, Japan, and PRESTO, Japan Science and Technology Agency (JST), 4-1-8 Honcho Kawaguchi, Saitama 332-0012, Japan}


\date{\today}

\begin{abstract}
Nonlinear oscillators have been used to model neurons that fire 
periodically in the absence of input. These oscillators, which are 
called neuronal oscillator, share some common response structures 
with other biological oscillations such as cardiac cells. In this study, 
we analyze the dependence of the global dynamics of an
impulse-driven stochastic neuronal oscillator on the relaxation rate
to the limit cycle, the strength of the intrinsic noise, and the
impulsive input parameters. To do this, we use a Markov operator
that both reflects the density evolution of the oscillator and is an
extension of the phase transition curve, which describes the phase
shift due to a single isolated impulse. Previously, we derived the
Markov operator for the finite relaxation rate that describes the
dynamics of the entire phase plane. Here, we construct a Markov
operator for the infinite relaxation rate that describes the
stochastic dynamics restricted to the limit cycle. In both cases,
the response of the stochastic neuronal oscillator to time-varying
impulses is described by a product of Markov operators. Furthermore,
we calculate the number of spikes between two consecutive impulses
to relate the dynamics of the oscillator to the number of spikes per
unit time and the interspike interval density. Specifically, we
analyze the dynamics of the number of spikes per unit time based on
the properties of the Markov operators. Each Markov operator can be
decomposed into stationary and transient components based on the
properties of the eigenvalues and eigenfunctions. This allows us to
evaluate the difference in the number of spikes per unit time
between the stationary and transient responses of the oscillator,
which we show to be based on the dependence of the oscillator on
past activity. Our analysis shows how the duration of the past neuronal 
activity depends on the relaxation rate, the noise strength and the 
impulsive input parameters.
\end{abstract}


\maketitle

\section{Introduction}
In nervous systems, information is transmitted via spikes; 
however, the question of whether information is carried via 
detailed spike patterns (temporal or timing coding) or simply by the 
number of spikes in a given time period (rate coding) is a 
subject of active debate \cite{RiekeWarlandSteveninckBialek1999}.
Each neuron in a neural network receives inputs from other neurons 
or outside of the neural network and transforms the inputs 
into spikes based on the intrinsic dynamics of the neuron. In the theory of 
artificial neural networks, the inputs are transformed by a function (for example, 
sigmoid function) and information carrier in an artificial neural network depends on 
the selection of the function. Thus, it is important to investigate how each neuron 
transforms the inputs into spikes. 
One necessary condition for temporal coding is that the spike 
generation of a neuron must not depend substantially on the past 
spike generation. Thus, the duration of the transient regime in neuronal activity
 must be short enough to achieve this independence. 
This means that the properties of the transient regime of the neuronal activity
may then offer insights into whether the pattern of spikes is a
possible information carrier in nervous systems. Because information
processing in nervous systems may occur in the transient regime, the
transient dynamics of neurons and neuronal models 
\cite{SegundoStiberAltshulerVibert1994,StiberIeongSegundo1997,YamanobePakdamanNomuraSato1998,Yamanobe2011},
and those of neural networks 
\cite{RabinovichBaronaSelverstonAbarbanel2006,RabinovichHuertaLaurent2010},
are the focus of analysis.

Nonlinear oscillators have been used to model neurons that fire periodically in 
the absence of input \cite{HoppensteadtIzhikevich1997,Izhikevich2010}. These 
nonlinear oscillators are called neuronal oscillator and are a subclass of nonlinear 
oscillators that are also found in a wide variety of biological and complex systems such as
cardiac cells \cite{GuevaraGlassShrier1981,GuevaraShrierGlassPerez1983},
respiratory rhythm generation 
\cite{SmithEllenbergerBallanyiRichterFeldman1991,Johnson1994},
Josephson junctions \cite{WiesenfeldColetStrogatz1998}, and climate
dynamics \cite{Jin1997}. In biological systems, the system may converge 
slowly to the asymptotic dynamics. In this case, the transient dynamics 
of the system can occur far from the asymptotic structure. Thus, 
it is important to analyze the global dynamics of the nonlinear oscillators 
to understand the transient dynamics.
However, as the global dynamics are difficult to understand,
higher-order dynamics are usually analyzed after a reduction to
lower-order dynamics. The most successful approach has been the
phase reduction method \cite{Kuramoto2003,Winfree2010}. If a
nonlinear oscillator is weakly perturbed, the trajectories will be
in the neighborhood of the limit cycle of the nonlinear oscillator.
Thus, the dynamics can be approximated by a vector field on the
limit cycle, and this makes it possible to represent the limit-cycle
dynamics in a higher-dimensional phase space by a one-dimensional
(1D) variable called the phase.

The impulse-driven nonlinear oscillators are often
used in the analysis of neuronal or biological oscillators
\cite{Winfree2010,Hoppensteadt1982,GlassMackey1988,GlassSun1994,StiberPakdamanVibertBoussardSegundoNomuraSatoDoi1997,
NomuraSatoDoiSegundoStiber1994,YamanobePakdamanNomuraSato1998,YamanobePakdaman2002,
PakdamanMestivier2001,CroisierGuevaraDauby2009}, and depending 
on the relaxation rate to the
limit cycle, the state points of the nonlinear oscillator can be far
from the limit cycle. For example, Glass and Sun \cite{GlassSun1994} analyzed the
dependence of the bifurcation structure of an impulse-driven
nonlinear oscillator on the relaxation rate. For their analysis,
they used a 2D extension of the phase transition curve for the
nonlinear oscillator, which represents the phase shift due to a 
single isolated impulse. 
However, it is usually only 
the dynamics in the asymptotic regime that are analyzed since 
there are few available tools for analyzing the dynamics in the 
transient regime.

Noise will also affect the dynamics, since intrinsic noise, e.g.,
ion channel noise \cite{WhiteRubinsteinKay2000}, might
restrict the accuracy of the spike generation. In general, nonlinear
systems are often influenced by stochastic fluctuations. In our
previous study, we introduced a Markov operator for an
impulse-driven stochastic neuronal oscillator that can approximate
the density evolution in the entire phase space of the oscillator
driven by time-varying impulses \cite{Yamanobe2011}.

In this paper, we analyze the global dynamics of a stochastic
neuronal oscillator driven by time-varying impulses 
by changing the relaxation rate to the limit
cycle, intrinsic noise strength, and input impulse parameters. We
introduce a Markov operator for an infinite relaxation rate using
the small disturbance asymptotic theory
\cite{KunitomoTakahashi2003,TakahashiTakeharaToda2011}, and this
operator describes the stochastic dynamics around the limit cycle.
We investigate the dynamics of the entire phase space without input
impulses using our Markov operator for finite and infinite
relaxation rates and analyze the response of the stochastic neuronal
oscillator to impulsive inputs by examining the effects of the
relaxation rate, intrinsic noise strength, and input impulse
parameters. For both the finite and infinite cases, the response of
the stochastic neuronal oscillator to time-varying impulses is
described by the product of the Markov operators. We can decompose
the Markov operator into stationary and transient components based
on the properties of eigenvalues and eigenfunctions to identify the
components that affect the current response. Moreover, we introduce
a stochastic rotation number to relate the dynamics of the
oscillator to the number of spikes per unit time and the interspike
interval (ISI) density to understand the steady state dynamics of
the oscillator. Specifically, we analyze the components of the
stochastic rotation number based on the properties of the Markov
operator. In relation to the information carrier in nervous systems,
we show how the past activity of the stochastic neuronal oscillator
affects the current firing rate. We demonstrate that there can exist the 
long-range dependence of the current neuronal activity on 
the past activity depending on the relaxation rate, the noise strength, 
input parameters.

\section{Methods}
\subsection{Stochastic Poincar\'{e} oscillator}
The Poincar\'{e} oscillator and its variant are a member of a set of systems that are 
widely used in the analysis of neuronal or biological oscillators
\cite{Winfree2010,Hoppensteadt1982,Guevara1982,Keener1984,GlassMackey1988,GlassSun1994,StiberPakdamanVibertBoussardSegundoNomuraSatoDoi1997,
NomuraSatoDoiSegundoStiber1994,YamanobePakdamanNomuraSato1998,Glass2001,YamanobePakdaman2002,
PakdamanMestivier2001,Izhikevich2010}. Based on \cite{GlassSun1994,Yamanobe2011},
we introduce the Poincar\'{e} oscillator here and summarize its
properties. The oscillator can be described in polar coordinates as
\begin{eqnarray}
\frac{dR_t^{(0)}}{dt}&=&KR_t^{(0)}(1-R_t^{(0)}) \nonumber \\
\frac{d\Phi_t^{(0)}}{dt}&=&1, \label{eq:RicPolar}
\end{eqnarray}
where $R_t^{(0)}\in \{x; x>0, x \in \mathbf{R}\}$ is the radial
coordinate, and $\Phi_t^{(0)} \in S^1$ ($S^1$ is the unit circle) is
the normalized angular coordinate that varies in [0,1); the
superscript (0) indicates the deterministic case and the subscript
represents the time \textit{t}. The positive parameter \textit{K} is
the relaxation rate to the limit cycle; the stable limit cycle is
the unit circle with period 1. Trajectories starting from any
initial point in the phase plane, except the origin, wind
counterclockwise around the origin and converge to the limit cycle
as $t \rightarrow \infty$. We define
$X_t^{(0)}=R_t^{(0)}\cos(2\pi\Phi_t^{(0)})$ as the ``membrane
potential" and $Y_t^{(0)}=R_t^{(0)}\sin(2\pi\Phi_t^{(0)})$ as the
``refractoriness". The spike occurs when the state point crosses the
positive $x$-axis. We consider the relationship between the state
point just before the \textit{n}th impulse and that just before the
(\textit{n} + 1)th impulse. Following \cite{GlassSun1994}, we define
the \textit{n}th impulsive stimulation by an instantaneous
horizontal shift by an amount $A_n$, where $n$ denotes the $n$th
impulse. If an impulse with amplitude $A_n$ shifts a state point
$(r_n,\phi_n)$  to the point $(r_n',\phi_n')$, the relation becomes
\begin{eqnarray}
r_n'&=&F_R(r_n,\phi_n)  = [r_n^2 + A_n^2 + 2 A_n r_n \cos(2\pi\phi_n) ]^{1/2} \nonumber \\
\phi_n'&=&F_\Phi(r_n,\phi_n) = \frac{1}{2\pi} \arccos \frac{r_n \cos(2 \pi \phi_n) + A_n}{F_R(r_n,\phi_n)}, \label{eq:TwoPTC}
\end{eqnarray}
where the subscripts $R$ and $\Phi$ denote the shift in the
directions of the radial and normalized angular coordinates,
respectively. To evaluate the arc-cosine function, we should take
$0<\phi_n'<0.5$ for $0<\phi_n<0.5$ and $0.5<\phi_n'<1$ for
$0.5<\phi_n<1$. Equation (\ref{eq:TwoPTC}) is the 2D version of the
phase transition curve of this oscillator, which represents the
phase shift due to a single isolated impulse.

After the \textit{n}th interimpulse interval $I_n$, the state point
starting from the initial point $(r_n',\phi_n')$, as determined by
Eq.~(\ref{eq:RicPolar}), is expressed as follows: 
\begin{eqnarray}
r_{n+1} &=& R_{I_n}^{(0)} = r_n'/\{(1-r_n')e^{-K I_n} + r_n'\} \nonumber \\
\phi_{n+1} &=& \Phi_{I_n}^{(0)} = \phi_n' + I_n \pmod{1}. \label{eq:ImpulseRes}
\end{eqnarray}

In the case of $K\rightarrow \infty$, the dynamics of the oscillator
are described only by the normalized angular coordinate. Thus, the
effect of the \textit{n}th impulse is 
\begin{eqnarray}
\phi_n'=\tilde{F}_{\Phi}(\phi_n) = \frac{1}{2\pi} \arccos
\frac{\cos(2 \pi \phi_n) +
A_n}{\sqrt{1+A_n^2+2A_n\cos(2\pi\phi_n)}}, \label{eq:1DPTC}
\end{eqnarray}
which defines the phase transition curve for this model. 
$\tilde{F}_{\Phi}(\phi_n)-\phi_n$ 
corresponds to phase response curve, which shows the 
phase shift due to an impulse, and the shape of the phase 
response curve is biphasic if $|A|<1$ \cite{Izhikevich2010}. 
The biphasic phase response curves are observed experimentally 
(for example, \cite{Tateno2007a}). The tilde denotes 
the function as $K \rightarrow \infty$. In what follows, we
use the tilde for functions and variables when it is necessary to
indicate that $K \rightarrow \infty$. Using this phase transition
curve, the state point just before the (\textit{n} + 1)th impulse
becomes 
\begin{eqnarray}
\phi_{n+1} = \tilde{\Phi}_{I_n}^{(0)} =  \phi_n'+I_n \pmod{1}, \label{eq:Expand1DPTC}
\end{eqnarray}
where $\phi_n' = \tilde{F}_{\Phi}(\phi_n)$. 
In previous study, we transform
Eq.~(\ref{eq:RicPolar}) into Cartesian coordinates and include a
noise term in the expression for the membrane potential.
In polar coordinates, the Poincar\'{e} oscillator with the noise term can be
expressed as \cite{Yamanobe2011}
\begin{eqnarray}
dR_t^{(\epsilon)} &=& K R_t^{(\epsilon)}(1-R_t^{(\epsilon )})dt +\frac{\epsilon^2}{2} \frac{\sin^2 (2\pi \Phi_t^{(\epsilon)})}{R_t^{(\epsilon)}}dt + \epsilon \cos (2 \pi \Phi_t^{(\epsilon)}) dW_t \nonumber \label{eq:RicPolardiff} \\
d\Phi_t^{(\epsilon)} &=& dt+\frac{\epsilon^2}{4\pi}\frac{\sin (4\pi
\Phi_t^{(\epsilon)})}{{R_t^{(\epsilon)}}^2}dt - \frac{\epsilon}{2
\pi} \frac{\sin (2 \pi \Phi_t^{(\epsilon)})}{R_t^{(\epsilon)}}dW_t.
\end{eqnarray}
where the superscript indicates the dependence of the random
variables on the strength of the noise term $\epsilon \in  (0,1]$,
for which $\epsilon=0$ gives the deterministic case. The 1D standard
Wiener process is denoted by $W_t$. We refer to
Eq.~(\ref{eq:RicPolardiff}) as a stochastic Poincar\'{e} oscillator.

\subsection{Stochastic phase transition operator}
In our previous study \cite{Yamanobe2011}, we introduced a Markov
operator that relates the density of the state points just before
the \textit{n}th impulse to that just before the (\textit{n} + 1)th
impulse. We called this Markov operator a stochastic phase
transition operator (SPTO). In what follows, we derive SPTOs for
infinite \textit{K}. According to Eq.~(\ref{eq:TwoPTC}), the state
point just before the \textit{n}th impulse $(r_n,\phi_n)$ shifts to
$(r_n',\phi_n')$ after the \textit{n}th impulse. After this shift,
the dynamics of the state point are described by the following 
integral equation with the initial condition
$(r_n',\phi_n')$ defined by Eq.~(\ref{eq:TwoPTC}): 
\begin{eqnarray}
R_{I_n}^{(\epsilon)} &=& r_n' + K \int_0^{I_n} R_s^{(\epsilon)}(1-R_s^{(\epsilon)})ds + \frac{\epsilon^2}{2} \int_0^{I_n}\frac{\sin^2 (2\pi \Phi_s^{(\epsilon)})}{R_s^{(\epsilon)}}ds + \epsilon\int_0^{I_n} \cos (2 \pi \Phi_s^{(\epsilon)}) dW_s \nonumber \\
\Phi_{I_n}^{(\epsilon)}&=& \phi_n' + I_n +
\frac{\epsilon^2}{4\pi}\int_0^{I_n}\frac{\sin (4\pi
\Phi_s^{(\epsilon)})}{{R_s^{(\epsilon)}}^2}ds - \frac{\epsilon}{2
\pi}\int_0^{I_n} \frac{\sin (2 \pi
\Phi_s^{(\epsilon)})}{R_s^{(\epsilon)}}dW_s,\pmod{1},
\label{eq:FullRicPolarMod}
\end{eqnarray}
In the case of $K\rightarrow\infty$, the dynamics of the stochastic
Poincar\'{e} oscillator can be described by the dynamics on the
limit cycle. As $K \rightarrow \infty$, $R_{I_n}^{(\epsilon)}=1$ and
the dynamics of $\Phi_t^{(\epsilon)}$ are given by
\begin{equation}
\Phi_t^{(\epsilon)}=\tilde{\Phi}_{I_n}^{(\epsilon)}= \phi_n' + I_n + \frac{\epsilon^2}{4\pi}\int_0^{I_n} \sin (4\pi \tilde{\Phi}_s^{(\epsilon)})ds - \frac{\epsilon}{2 \pi}\int_0^{I_n} \sin (2 \pi \tilde{\Phi}_s^{(\epsilon)})dW_s, \pmod{1}, \label{eq:RicPhaseMod}
\end{equation}
where $\phi_n'$ is defined by Eq.~(\ref{eq:1DPTC}). We call
Eq.~(\ref{eq:RicPhaseMod}) a phase equation and note that it
includes the modification term suggested by Yoshimura and Arai
\cite{YoshimuraArai2008} since we take $K \rightarrow \infty$ after
the coordinate transform.

We introduce a new random variable
$\Theta_t^{(\epsilon)}=\Phi_t^{(\epsilon)} \pmod{1}$ that takes a
value in $\mathbf{R}$ and explicitly indicates the rotation around
the origin. Equation (\ref{eq:FullRicPolarMod}) then becomes 
\begin{eqnarray}
R_{I_n}^{(\epsilon)} &=& r_n' + K \int_0^{I_n} R_s^{(\epsilon)}(1-R_s^{(\epsilon)})ds + \frac{\epsilon^2}{2} \int_0^{I_n}\frac{\sin^2 (2\pi \Theta_s^{(\epsilon)})}{R_s^{(\epsilon)}}ds + \epsilon\int_0^{I_n} \cos (2 \pi \Theta_s^{(\epsilon)}) dW_s  \nonumber \\
\Theta_{I_n}^{(\epsilon)}&=& \phi_n' + I_n + \frac{\epsilon^2}{4\pi}\int_0^{I_n}\frac{\sin (4\pi \Theta_s^{(\epsilon)})}{{R_s^{(\epsilon)}}^2}ds - \frac{\epsilon}{2 \pi}\int_0^{I_n} \frac{\sin (2 \pi \Theta_s^{(\epsilon)})}{R_s^{(\epsilon)}}dW_s. \label{eq:FullRicPolar}
\end{eqnarray}
Similarly, for Eq.~(\ref{eq:RicPhaseMod}), we introduce the random
variable
$\tilde{\Theta}_{I_n}^{(\epsilon)}=\tilde{\Phi}_{I_n}^{(\epsilon)}
\pmod{1}$, and Eq.~(\ref{eq:RicPhaseMod}) becomes 
\begin{equation}
\tilde{\Theta}_{I_n}^{(\epsilon)}= \phi_n' + I_n + \frac{\epsilon^2}{4\pi}\int_0^{I_n} \sin (4\pi \tilde{\Theta}_s^{(\epsilon)})ds - \frac{\epsilon}{2 \pi}\int_0^{I_n} \sin (2 \pi \tilde{\Theta}_s^{(\epsilon)})dW_s. \label{eq:RicPhase}
\end{equation}
In what follows, we call $\Theta_{I_n}^{(\epsilon)}$ and $\tilde{\Theta}_{I_n}^{(\epsilon)}$ a lifted angular
coordinate. To calculate the stochastic kernels of the SPTO, i.e.,
the transition density corresponding to a given stochastic
differential equation, we apply the small disturbance asymptotic
theory, which is an asymptotic expansion of the stochastic processes
\cite{KunitomoTakahashi2003, TakahashiTakeharaToda2011}. To apply
this theory, we assume that the diffusion coefficients in
Eqs.~(\ref{eq:FullRicPolar}) and (\ref{eq:RicPhase}) are not zero
for any $s>0$. This assumption guarantees the asymptotic expansion
of the transition density around the normal distribution density.
Thus, we have to apply the theory to Eqs.~(\ref{eq:FullRicPolar})
and (\ref{eq:RicPhase}) separately to calculate the transition
density. We derived the stochastic kernel for the full equation
(Eq.~(\ref{eq:FullRicPolar})) in our previous study
\cite{Yamanobe2011}. Here, we derive
the stochastic kernel of the SPTO for Eq.~(\ref{eq:RicPhase}). We
first expand $\tilde{\Theta}_{t}^{(\epsilon)}$ with respect to
$\epsilon$ as
\begin{equation}
\tilde{\Theta}_{I_n}^{(\epsilon)} = \tilde{\Theta}_{I_n}^{(0)}+\epsilon A_{1\tilde{\Theta} I_n}+o(\epsilon), \nonumber
\end{equation}
where $\tilde{\Theta}_{I_n}^{(0)}$ is a deterministic solution of
Eq.~(\ref{eq:RicPhase}) and $A_{1\tilde{\Theta} I_n} =
\frac{\partial \tilde{\Theta}_{I_n}^{(\epsilon)}}{\partial \epsilon}
\bigr |_{\epsilon=0}$, where the subscript $\tilde{\Theta}I_n$
denotes the derivative of $\tilde{\Theta}_{I_n}^{(\epsilon)}$
explicitly. 
The derivative is
\begin{eqnarray}
A_{1\tilde{\Theta} I_n}&=&-\frac{1}{2\pi}\int_0^{I_n} \sin(2\pi\tilde{\Theta}_s^{(0)}) dW_s. \nonumber
\end{eqnarray}
To consider the stochastic dynamics around the deterministic
solution $\tilde{\Theta}_{I_n}^{(0)}$, we introduce a new random
variable
$\tilde{S}_{I_n}^{(\epsilon)}=(\tilde{\Theta}_{I_n}^{(\epsilon)}-\tilde{\Theta}_{I_n}^{(0)})/\epsilon$.
The expansion of $\tilde{S}_{I_n}^{(\epsilon)}$ with respect to
$\epsilon$ gives 
\begin{eqnarray}
\tilde{S}_{I_n}^{(\epsilon)} = A_{1\tilde{\Theta} I_n} + o(1).
\nonumber
\end{eqnarray}
The asymptotic expansion of the characteristic function of
$\tilde{S}_{I_n}^{(\epsilon)}$ with respect to $\epsilon$ is 
\begin{eqnarray}
\Psi(\mathbf{\xi}) &=& \mathbf{E}[\exp \{i \xi (A_{1\tilde{\Theta} I_n} + o(1))\}] \nonumber \\
&=& \mathbf{E}[ \exp (i\xi A_{1 \tilde{\Theta} I_n}) \{1+o(1)\}] \nonumber \\
&=&\mathbf{E}[\exp(i\xi A_{1\tilde{\Theta} I_n})] +o(1) \nonumber \\
&=& \exp\bigl\{-\frac{1}{2}(\Sigma_{\tilde{\Theta}I_n}(\phi_n)\xi^2)\bigr\}
+o(1),  \label{eq:CharReduced}
\end{eqnarray}
where $\xi \in \mathbf{R}$ and $\Sigma_{\tilde{\Theta}I_n}(\phi_n)
=\mathbf{E}[A_{1\tilde{\Theta} I_n}^2]
=\Bigl(\frac{1}{2\pi}\Bigr)^3\bigl\{\pi I_n -\frac{1}{2}\cos(2\pi
(2\tilde{F}_{\Phi}(\phi_n)+I_n))\sin(2\pi I_n)\bigr\}$. In
Eq.~(\ref{eq:CharReduced}), the second equality is derived from the
expansion of the exponential function with respect to $\epsilon$.
The fourth equality is derived from the Gaussianity of
$A_{1\tilde{\Theta} I_n}$. Using the inverse Fourier transform of
Eq.~(\ref{eq:CharReduced}), we obtain 
\begin{eqnarray}
f_{\tilde{S}_{I_n}^{(\epsilon)}}(s)&=&
n[s;0,\Sigma_{\tilde{\Theta}I_n}(\phi_n)]+o(1),  \label{eq:DensityPhase}
\end{eqnarray}
where $s \in \mathbf{R}$ and
$n[s;0,\Sigma_{\tilde{\Theta}I_n}(\phi_n)]$ is the density of a 1D
Gaussian distribution with zero mean and a variance of
$\Sigma_{\tilde{\Theta}I_n}(\phi_n)$; the subscript denotes the
density of the random variable $\tilde{S}_{I_n}^{(\epsilon)}$.
Equation (\ref{eq:DensityPhase}) is derived by expanding around the
solution $\tilde{\Theta}_{I_n}^{(0)}$, and the obtained density is
the 1D Gaussian distribution in the direction of $\tilde{\Theta}$.
We use Eq.~(\ref{eq:DensityPhase}) to approximate the density of
$\tilde{\Theta}_{I_n}^{(\epsilon)}$ as 
\begin{eqnarray}
f_{{\tilde{\Theta}}_{I_n}^{(\epsilon)}}(\theta;\phi_n) &\sim& n[\theta;\tilde{\Theta}_{I_n}^{(0)}(\phi_n),\epsilon^2\Sigma_{\tilde{\Theta}I_n}(\phi_n)], \label{eq:DensityKernelPhase}
\end{eqnarray}
where $s=(\theta-\Theta_{I_n}^{(0)})/\epsilon$. The dependence on
$\phi_n$ is explicit in the term
$\tilde{\Theta}_{I_n}^{(0)}(\phi_n)$, and the function
$n[\theta;\tilde{\Theta}_{I_n}^{(0)}(\phi_n),\epsilon^2\Sigma_{\tilde{\Theta}_{I_n}}(\phi_n)]$
depends on $\phi_n$ via Eqs.~(\ref{eq:1DPTC}) and
(\ref{eq:Expand1DPTC}). In this way, the difficulty of the
discontinuity caused by the impulse is avoided when approximating
$f_{{\tilde{\Theta}}_{I_n}^{(\epsilon)}}(\theta;\phi_n)$. Using
Eq.~(\ref{eq:DensityKernelPhase}), the stochastic kernel for the
phase equation is given by
\begin{eqnarray}
g_{\infty,\epsilon,A_n,I_n}(\phi;\phi_n) = \sum_{p=-\infty}^{p=+\infty}n[\phi+p;\Theta_{I_n}^{(0)}(\phi_n),\epsilon^2\Sigma_{\tilde{\Theta}_{I_n}}(\phi_n)], \label{eq:ReducedTranProbDens}
\end{eqnarray}
where $\phi=\theta \pmod{1}$. The summation with respect to $\phi$
takes into account multiple rotations around the origin, and the
dependence of the stochastic kernel on $K \rightarrow \infty$, $\epsilon$, $A_n$, and $I_n$
is denoted explicitly using the subscript.

Using this stochastic kernel, the evolution of the density just
before the \textit{n}th impulse to that just before the (\textit{n}
+ 1)th impulse is determined by
\begin{eqnarray}
&&h_{n+1}(\phi) = \int_0^1 g_{\infty,\epsilon,A_n,I_n}(\phi;\phi_n) h_{n}(\phi_n) d\phi_n  = \mathbf{P}_{\infty,\epsilon,A_n,I_n}h_n(\phi), \label{eq:ReducedMarkovOperator}
\end{eqnarray}
where $h_n$ is the density of the phase equation just before the
\textit{n}th impulse. We call the $\mathbf{P}_{\infty,\epsilon,A,I}$
operator a 1D-SPTO.

Using the stochastic kernel for the full equation (see
Eq.~(8) in \cite{Yamanobe2011}), the SPTO that expresses
the relationship between the density just before the \textit{n}th
impulse and that just before the (\textit{n} + 1)th impulse is
written as 
\begin{eqnarray}
h_{n+1}(r,\phi)=\int_{0}^{1} \int_{0}^{\infty} g_{K,\epsilon,A_n,I_n}(r,\phi;r_n,\phi_n) h_{n}(r_n,\phi_n)dr_n d\phi_n = \mathbf{P}_{K,\epsilon,A_n,I_n}h_n(r,\phi), \label{eq:FullMarkovOperator}
\end{eqnarray}
where $h_n$ is the density just before the \textit{n}th impulse for
the full equation. It should be noted that we use $h$ as the density
for both the full and phase equations. When these densities need to be 
distinguished, we show the variables of $h$ explicitly. The
$\mathbf{P}_{K,\epsilon,A,I}$ operator is referred to as a 2D-SPTO,
which is a 2D generalization of the phase transition curve with a
stochastic term. In what follows, we use
$\mathbf{P}_{K,\epsilon,A,I}$ for $K \in (0,\infty]$.

\subsection{Spectral properties of the SPTO}

The SPTO is a linear operator and contains all the information about
the density evolution. In what follows, we discretize the SPTO to
analyze its properties. We use numerical integration to approximate
the integral equations, Eqs.~(\ref{eq:ReducedMarkovOperator}) and
(\ref{eq:FullMarkovOperator}). In the case of
Eq.~(\ref{eq:FullMarkovOperator}), we first truncate the integration
range of the \textit{r}-axis, since the density in the direction of
\textit{r} decreases rapidly to zero as \textit{r} increases. The
integration range for $r_n$ is large enough to approximate the
integral equation as 
\begin{eqnarray}
h_{n+1}(r,\phi) \approx \int_0^1 \int_0^{a} g_{K,\epsilon,A_n,
I_n}(r,\phi;r_n,\phi_n)h_n(r_n,\phi_n)dr_n d\phi_n.
\label{eq:TrancatedFullMarkovOperator}
\end{eqnarray}
We then discretize Eq.~(\ref{eq:TrancatedFullMarkovOperator}) using
quadrature rules: 
\begin{eqnarray}
h_{n+1}(r_k,\phi_l) \approx \sum_{j=1}^{N_{\phi_n}} \sum_{i=1}^{N_{r_n}} w_i^{r_n} w_j^{\phi_n}  g_{K,\epsilon,A_n, I_n}(r_k,\phi_l;r_{n,i},\phi_{n,j})h_n(r_{n,i},\phi_{n,j}).
\label{eq:DiscretizedFullMarkovOperator}
\end{eqnarray}
Legendre--Gauss quadrature for the integration with respect to $r_n$
and the trapezoidal rule for the integration with respect to
$\phi_n$ are used; the same nodes are used for the approximation,
i.e., the pair $(r_n,\phi_n)$ and $(r,\phi)$ have the same nodes.

The numerical calculations were performed with Matlab. To calculate
the density evolution, we set the discretization of the density
$h(r,\phi)$ as follows: 

\begin{equation}
	\begin{pmatrix}
	h(r_1,\phi_1) &\dots & h(r_1,\phi_{N_{\phi_n}}) \\
	\vdots & \ddots & \vdots \\
	h(r_{N_{r_n}},\phi_1) & \dots & h(r_{N_{r_n}},\phi_{N_{\phi_n}}) \label{eq:DiscreDens}
	\end{pmatrix}.
\end{equation}
We transform this matrix as 
\begin{equation}
\begin{pmatrix}
h(r_1,\phi_1) \\
\vdots \\
h(r_{N_{r_n}},\phi_1) \\
\vdots \\
h(r_1,\phi_{N_{\phi_n}}) \\
\vdots \\
h(r_{N_{r_n}},\phi_{N_{\phi_n}})
\end{pmatrix}.
\label{eq:DiscreH}
\end{equation}
To calculate the density evolution using Eq.~(\ref{eq:DiscreH}), we
constructed the corresponding stochastic matrix given as 
\begin{equation}
\begin{pmatrix}
g_{K,\epsilon,A,I1,1,1,1} & \dots & g_{K,\epsilon,A,I1,1,N_{r_n},1} & \dots &g_{K,\epsilon,A,I1,1,1,N_{\phi_n}}  & \dots & g_{K,\epsilon,A,I1,1,N_{r_n},N_{\phi_n}}\\
\vdots & \ddots & \vdots & \ddots & \vdots & \ddots & \vdots \\
g_{K,\epsilon,A,IN_{r_n},1,1,1}  & \dots &   g_{K,\epsilon,A,IN_{r_n},1,N_{r_n},1} & \dots & g_{K,\epsilon,A,IN_{r_n},1,1,N_{\phi_n}}& \dots & g_{K,\epsilon,A,IN_{r_n},1,N_{r_n},N_{\phi_n}}  \\
\vdots & \ddots & \vdots & \ddots & \vdots & \ddots & \vdots \\
g_{K,\epsilon,A,I1,N_{\phi_n},1,1} & \dots &g_{K,\epsilon,A,I1,N_{\phi_n},N_{r_n},1}  & \dots & g_{K,\epsilon,A,I1,N_{\phi_n},1,N_{\phi_n}}  & \dots &  g_{K,\epsilon,A,I1,N_{\phi_n},N_{r_n},N_{\phi_n}}\\
\vdots & \ddots & \vdots & \ddots & \vdots & \ddots & \vdots \\
g_{K,\epsilon,A,IN_{r_n},N_{\phi_n},1,1} & \dots &  g_{K,\epsilon,A,IN_{r_n},N_{\phi_n},N_{r_n},1} & \dots & g_{K,\epsilon,A,IN_{r_n},N_{\phi_n},1,N_{\phi_n}} & \dots & g_{K,\epsilon,A,IN_{r_n},N_{\phi_n},N_{r_n},N_{\phi_n}}
\end{pmatrix}, \label{eq:DiscreNumFullOpe}
\end{equation}
where we set $g_{K,\epsilon,A,I,i,j,k,l} =
g_{K,\epsilon,A,I}(\nu_i,\phi_j;r_{k},\phi_{l})w_k^{r} w_l^{\phi}$
for a concise representation of the matrix. Usually, the stochastic
matrix is defined as a square matrix in which each row consists of
nonnegative real numbers that sum to 1. However, for convenience, we
set the stochastic matrix, which is a discretization of the SPTO, to
a square matrix whose columns consist of nonnegative real numbers
and for which each column sums to 1.

Similarly, we approximate Eq.~(\ref{eq:ReducedMarkovOperator}) using
the  trapezoidal rule as 
\begin{eqnarray}
h_{n+1}(\phi_k) \approx \sum_{j=1}^{N_{\phi_n}} w_j^{\phi_n}  g_{\infty,\epsilon,A_n, I_n}(\phi_k;\phi_{n,j})h_n(\phi_{n,j}). \label{eq:ReducedMarkovOperatorTrapz}
\end{eqnarray}
In this case, the representations of the matrix and vectors are
easily deduced from Eq.~(\ref{eq:ReducedMarkovOperatorTrapz}).

Let us fix the input parameters \textit{A} and \textit{I} of the
SPTO. We analyze the spectral properties of the discretized SPTO
because the dynamics of the SPTO matrix are determined by the
eigenvalues and eigenfunctions (eigenvectors). Let $\{ \alpha_{i}
\}$ and $\{ e_i \}$ be the eigenvalues of the discretized SPTO,
sorted in descending order according to their moduli, and the
corresponding eigenfunctions, respectively ($i=1,2,...,N_{\phi_n}$
or $N_{\phi_n}N_{r_n}$, where $N_{\phi_n}$ and $N_{\phi_n}N_{r_n}$
are the dimensions of the discretized SPTOs in
Eqs.~(\ref{eq:ReducedMarkovOperator}) and
(\ref{eq:FullMarkovOperator}), respectively). Since the stochastic
kernel is positive and the discretized SPTO is a positive stochastic
matrix, the properties of the matrix can be summarized as follows
\cite{Gantmacher1959}:

\begin{description}
\item[I.] $\alpha_{1}=1$ and has a multiplicity of one. The corresponding eigenfunction has a unique invariant density $h^{*}_{K,\epsilon,A,I}$ or is $e_1$ with positive coordinates, i.e., the discretized SPTO is ergodic.
\item[II.] $|\alpha_{i}|<1$ for all eigenvalues different from $1$.
\end{description}

Hence, the eigenvalues $\alpha_{i}$ and eigenfunctions $e_{i}$ with
$i\geqq2$ have transient dynamics or contain the ``dynamic"
information of the discretized SPTO, whereas the invariant density
$h^{*}_{K,\epsilon,A,I}$ or $e_{1}$ has stationary dynamics or ``static"
information of the discretized SPTO. In other words, the invariant
density shows the response of the oscillator to periodic impulses as
time goes to infinity. Based on these properties, the discretized
SPTO is decomposed into two parts 
\begin{equation}
    \mathbf{P}_{K,\epsilon,A,I}=\mathbf{V}_{K,\epsilon,A,I} + \mathbf{Q}_{K,\epsilon,A,I}, \label{eq:SpecDecomp}
\end{equation}
where $\mathbf{V}_{K,\epsilon,A,I}$ represents the stationary
dynamics, i.e., for a density \textit{h},
$\mathbf{V}_{K,\epsilon,A,I} h=h_{K,\epsilon,A,I}^*$, and
$\mathbf{Q}_{K,\epsilon,A,I}$ corresponds to the transient dynamics.
It should be noted that the spectral decomposition in
Eq.~(\ref{eq:SpecDecomp}) also holds for the ``original" SPTO, since
the SPTO is a constrictive Markov operator. The constrictiveness
means that $\mathbf{P}_{K,\epsilon,A,I}^n h$ does not concentrate on
a set of very small or vanishing measures as $n \rightarrow \infty$
(see \cite{LasotaMackey1998, DingZhou2009} for an explanation of
constrictiveness and Proposition 5.3.2 in \cite{LasotaMackey1998} to
verify the constrictiveness of the SPTO). In short, since the
stochastic kernel of the SPTO is positive, the SPTO is
asymptotically stable \cite{LasotaMackey1998} and thus constrictive.

For Eq.~(\ref{eq:FullMarkovOperator}), the discretization of
$\mathbf{V}_{K,\epsilon,A,I}$ in Eq.~(\ref{eq:SpecDecomp}) is
express as follows: 
\begin{equation}
\mathbf{h}_{K,\epsilon,A,I}^* \mathbf{1}^T, \label{eq:DiscreInvOpe}
\end{equation}
where \textit{T} denotes the transpose, {\bf{1}} represents an
$(N_{r_n} \times N_{\phi_n})$ vector of 1's, and
$\mathbf{h}_{K,\epsilon,A,I}^*$ is 
\begin{equation}
\begin{pmatrix}
h_{K,\epsilon,A,I}^*(r_1,\phi_1) \\
\vdots \\
h_{K,\epsilon,A,I}^*(r_{N_{r_n}},\phi_1) \\
\vdots \\
h_{K,\epsilon,A,I}^*(r_1,\phi_{N_{\phi_n}}) \\
\vdots \\
h_{K,\epsilon,A,I}^*(r_{N_{r_n}},\phi_{N_{\phi_n}})
\end{pmatrix}.
\end{equation}

\subsection{Stochastic phase locking}
Let us set the eigenvalues of the discretized SPTO $\alpha_i =
\rho_i \exp(2\pi j\kappa_i)$, with \textit{j} as the imaginary unit
and where $\rho_i$ and $\kappa_i$ are the modulus and angle of
$\alpha_i$, respectively, and the corresponding eigenfunctions $e_i$
to fixed values. Applying the discretized
$\mathbf{P}_{K,\epsilon,A,I}$ to $e_i$ for a total of \textit{k}
times yields 
\begin{eqnarray}
\mathbf{P}_{K,\epsilon,A,I}^k e_i = \alpha_i^k e_i = \rho_i^k \exp(2\pi j k
\kappa_i) e_i \quad (k=1, 2, \cdots). \nonumber
\end{eqnarray}
Based on the dynamic information of the discretized
$\mathbf{P}_{K,\epsilon,A,I}$, that is, eigenvalues $\alpha_i$ and
eigenfunctions $e_i$ with $i \geq 2$, Doi et al.
\cite{DoiInoueKumagai1998} defined stochastic bifurcation as the
abrupt (not smooth) change of the eigenvalues of the operator from
complex to real values at a possible stochastic bifurcation point.
Furthermore, they also defined stochastic phase locking as the
response that satisfies the following condition in addition to the
stochastic bifurcation condition: In a certain range, there exists
an \textit{i} that satisfies
\begin{equation}
\mathbf{P}_{K,\epsilon,A,I}^p e_i = \alpha_i^p e_i = \rho_i^p \exp(2\pi j p
\kappa_i)e_i = \rho_i^p e_i \quad (i \geq 2). \nonumber
\end{equation}
In this study, we use the second eigenvalue of the operator to
define the stochastic bifurcation and stochastic phase locking for categorizing 
the dynamics of the stochastic Poincar\'{e} oscillator. 
The definition of the stochastic bifurcation is still in an active debate. 
For the detailed discussion of the stochastic bifurcation, please see 
\cite{TatenoDoiSatoRicciardi1995,DoiInoueKumagai1998,Tateno1998,InoueDoiKumagai2001,Tateno2002,Arnold2008,BorisyukRassoulAgha}.

\subsection{Contribution to the current state from the past states}

Let us consider a sequence of \textit{n} impulses and the
corresponding product of discretized SPTOs
$\mathbf{H}_{n,1}=\mathbf{P}_{K,\epsilon,A_n,I_n}\mathbf{P}_{K,\epsilon,A_{n-1},I_{n-1}}\cdots
\mathbf{P}_{K,\epsilon,A_1,I_1}$ that describes the response of the
stochastic Poincar\'{e} oscillator to the impulses. We set the
current state to the state just before the (\textit{n} + 1)th
impulse, which is represented by $h_{n+1}$($=\mathbf{H}_{n,1}h_1$).
Since the current density is determined by the product of the
discretized SPTOs, the current density depends on the past activity
of the stochastic Poincar\'{e} oscillator. The structure of the
products
$\mathbf{P}_{K,\epsilon,A_n,I_n}\mathbf{P}_{K,\epsilon,A_{n-1},I_{n-1}}\cdots
\mathbf{P}_{K,\epsilon,A_1,I_1}$ tells us how the past activity
affects the current density.

Using Eq.~(\ref{eq:SpecDecomp}), the product is expressed as follows
\cite{Yamanobe2011}: 
\begin{eqnarray}
h_{n+1} = \mathbf{H}_{n,1}h_1
&=& \Bigl \{\mathbf{V}_{K,\epsilon,A_n,I_n} + \sum_{i=1}^{n-1}\Bigl(\prod_{j=0}^{n-i-1} \mathbf{Q}_{K,\epsilon,A_{n-j}, I_{n-j}} \Bigr) \mathbf{V}_{K,\epsilon,A_i,I_i} + \prod_{l=0}^{n-1}\mathbf{Q}_{K,\epsilon,A_{n-l},I_{n-l}}\Bigr \}h_1 \nonumber \\
&=& h_{K,\epsilon,A_n,I_n}^* + \sum_{i=1}^{n-1} \Bigl(\prod_{j=0}^{n-i-1} \mathbf{Q}_{K,\epsilon,A_{n-j}, I_{n-j}} \Bigr) h_{K,\epsilon,A_i,I_i}^* + \prod_{l=0}^{n-1} \mathbf{Q}_{K,\epsilon,A_{n-l},I_{n-l}} h_1, \label{eq:DecompProduct}
\end{eqnarray}
where $h_{K,\epsilon,A_i,I_i}^{*}$ is the invariant density of
$\mathbf{P}_{K,\epsilon,A_i,I_i}$, $(1\leq i \leq n)$. It should be
noted that Eq.~(\ref{eq:DecompProduct}) holds for both ``original''
and ``discretized'' SPTOs since Eq.~(\ref{eq:SpecDecomp}) holds for
both. Equation (\ref{eq:DecompProduct}) suggests that the invariant
density at the last impulse, e.g., the invariant density of
$\mathbf{P}_{K,\epsilon,A_n,I_n}$, always appears in the equation of
the product as is. If all the transient components of each
discretized SPTO in the product are zero matrices, then the density
is always equal to the invariant density at the last impulse. The
second term in Eq.~(\ref{eq:DecompProduct}) shows the effect of the
difference between adjacent invariant densities, as the following
equation holds for each term in the second term:
\begin{eqnarray}
\mathbf{Q}_{K,\epsilon,A_{i},I_{i}}\mathbf{V}_{K,\epsilon,A_{i-1},I_{i-1}}h_1=\mathbf{Q}_{K,\epsilon,A_{i},I_{i}}h_{K,\epsilon,A_{i-1},I_{i-1}}^*=\mathbf{P}_{K,\epsilon,A_i,I_i}h_{K,\epsilon,A_{i-1},I_{i-1}}^* - h_{K,\epsilon,A_i,I_i}^*. \label{eq:SecondTermProduct}
\end{eqnarray}
This means that the contribution of this term becomes small if the
difference between adjacent invariant densities is small. The third
term in Eq.~(\ref{eq:DecompProduct}) describes the dependence on the
initial density.

\begin{figure}[htbp]
\begin{center}
\includegraphics[scale=1]{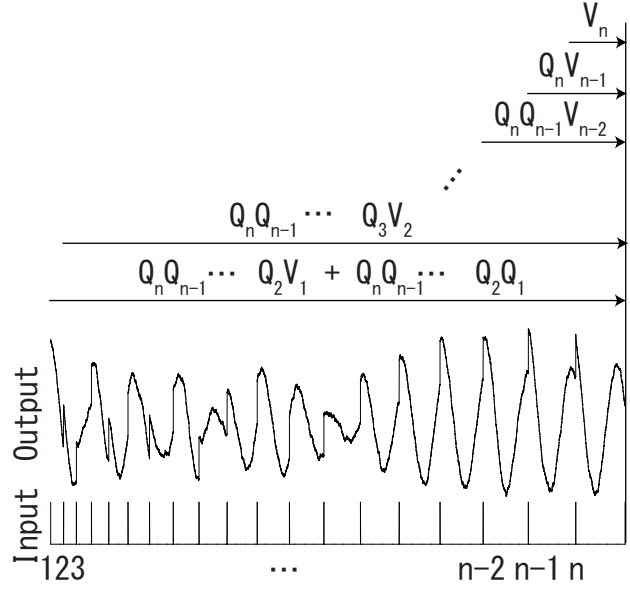}
\end{center}
\caption{ {\bf Schematic diagram explaining the dependence of the
current density on the past activity of the stochastic Poincar\'{e}
oscillator.} The terms of the product of operators on the right-hand
side of the second equation in Eq.~(\ref{eq:DecompProduct}) are
plotted as a function of the input impulse number. The dynamics of
the stochastic Poincar\'{e} oscillator are governed by the product
of the stochastic phase transition operators (SPTOs) (for a detailed
explanation of the SPTO, see Eqs.~(\ref{eq:ReducedMarkovOperator})
and (\ref{eq:FullMarkovOperator})), and the SPTO
$\mathbf{P}_{K,\epsilon,A_i,I_i}$ expresses the relationship between
the density just before the $i$th impulse to that just before the
$(i+1)$th impulse: $\mathbf{P}_{K,\epsilon,A_i,I_i} =
\mathbf{V}_{K,\epsilon,A_i,I_i}+ \mathbf{Q}_{K,\epsilon,A_i,I_i}$,
where $\mathbf{V}_{K,\epsilon,A_i,I_i}$ denotes the stationary
dynamics and $\mathbf{Q}_{K,\epsilon,A_i,I_i}$ represents the
transient dynamics. In this figure, the operators $\mathbf{V}_i$ and
$\mathbf{Q}_i$ denote $\mathbf{V}_{K,\epsilon,A_i,I_i}$ and
$\mathbf{Q}_{K,\epsilon,A_i,I_i}$, respectively. If all the
transient components of each $\mathbf{P}_{K,\epsilon,A_i,I_i}$ in
the product are zero operators, then the current density is
determined by the invariant component of the last impulse
$\mathbf{V}_n$. The ``output'' is the membrane potential of the
stochastic Poincar\'{e} oscillator, and the ``input'' shows the
input impulses added to the stochastic Poincar\'{e} oscillator. }
\label{Figure_label1}
\end{figure}

Figure \ref{Figure_label1} illustrates how each term in the second
and third terms of Eq.~(\ref{eq:DecompProduct}) affect the current
density. For example, $\mathbf{Q}_{K,\epsilon,A_n,I_n}
\mathbf{V}_{K,\epsilon,A_{n-1},I_{n-1}}$ is produced by
$\mathbf{P}_{K,\epsilon,A_n,I_n}\mathbf{P}_{K,\epsilon,A_{n-1},I_{n-1}}$,
since $\mathbf{V}_{K,\epsilon,A_i,I_i}$ and
$\mathbf{Q}_{K,\epsilon,A_i,I_i}$ are from
$\mathbf{P}_{K,\epsilon,A_i,I_i}$ $(1 \leq i \leq n)$, respectively.
In other words,
$\mathbf{Q}_{K,\epsilon,A_n,I_n}\mathbf{V}_{K,\epsilon,A_n,I_n}$ is
determined by the input parameters of the $n$th and $(n-1)$th
impulses. Thus, we treat the term  $\mathbf{Q}_{K,\epsilon,A_n,I_n}
\mathbf{V}_{K,\epsilon,A_{n-1},I_{n-1}}$ as the effect produced by
the (\textit{n} $-$ 1)th and \textit{n}th impulses as in Figure
\ref{Figure_label1}. In this way, we can attribute the past neuronal
activity of the stochastic Poincar\'{e} oscillator to the components
in Eq.~(\ref{eq:DecompProduct}). To evaluate the relative
contribution of each term that contains information about past
activity, we use the following 1-norm of a discretized operator
{\bf{A}}: 
\begin{eqnarray}
\|\mathbf{A}\|_1=\sup_{\mathbf{x}
\neq\mathbf{0}}\frac{\|\mathbf{A}\mathbf{x}\|_1}{\|\mathbf{x}\|_1} =
\max_{1\leq j \leq m}\sum_{i=1}^m|a_{ij}| w_i, \label{eq:1-norm}
\end{eqnarray}
where {\bf{x}} is a vector, and $\|\mathbf{x}\|_1=\sum_{i=1}^{m}
|x_i|w_i$ ($m$ is the dimension of $\mathbf{x}$ and $w_i$ is
determined by Eq.~(\ref{eq:ReducedMarkovOperatorTrapz}) for
$K=\infty$ and by Eqs.~(\ref{eq:DiscretizedFullMarkovOperator}),
(\ref{eq:DiscreDens}), (\ref{eq:DiscreH}), and
(\ref{eq:DiscreNumFullOpe}) for finite values of $K$, since the
trapezoidal rule and quadrature are used for the numerical
integration). Since the discretized SPTO is a positive matrix, the
product of discretized SPTOs is weakly ergodic
\cite{Seneta2007,IpsenSelee2011}. The weak ergodicity leads to the
following property for any densities $h$ and $h'$: 
\begin{eqnarray}
\|\mathbf{H}_{n,n_0}h-\mathbf{H}_{n,n_0}h'\|_1 \rightarrow 0 \quad
\mbox{for all } n_0, \mbox{as } n \rightarrow \infty,
\end{eqnarray}
where $\mathbf{H}_{n,n_0}=\mathbf{P}_{K,\epsilon,A_{n},I_{n}}
\mathbf{P}_{K,\epsilon,A_{n-1},I_{n-1}} \cdots
\mathbf{P}_{K,\epsilon,A_{n_0},I_{n_0}}$, and $n_0$ and $n$ are
positive integers with $n \geq n_0$. This means that the product of
discretized SPTOs loses its dependence on the initial density.
Because the third term in Eq.~(\ref{eq:DecompProduct}) is the only
term that depends on the initial density, the 1-norm of this term
goes to zero as $n \rightarrow \infty$.

\subsection{Stochastic rotation number}
To connect the density evolution and the firing rate of the
stochastic Poincar\'{e} oscillator, we calculate a stochastic
rotation number for the phase equation in Eq.~(\ref{eq:RicPhaseMod})
and the full equation in Eq.~(\ref{eq:FullRicPolarMod}) following
the definition in
\cite{YamanobePakdaman2002,NesseClarkBressloff2007}. Considering the
case for which the \textit{n}th impulse is added at $\phi_n$, the
lifted angular coordinate from Eq.~(\ref{eq:DensityKernelPhase})
just before the (\textit{n} + 1)th impulse is distributed as
follows: 
\begin{eqnarray}
n[\theta;\tilde{\Theta}_{I_n}^{(0)}(\phi_n),\epsilon^2 \Sigma_{\tilde{\Theta}I_n}(\phi_n)],
\end{eqnarray}
where it should be noted that $\theta \in \mathbf{R}$. That is,
$\theta$ includes multiple rotations around the origin. The mean
difference in the lifted angular coordinates of two consecutive
impulses becomes 
\begin{eqnarray}
w_{\infty,A_n, I_n}(\phi_n) &=& \int_{-\infty}^{\infty}(\theta-\phi_n) n[\theta;\tilde{\Theta}_{I_n}^{(0)}(\phi_n),\epsilon^2\Sigma_{\tilde{\Phi}I_n}(\phi_n)]d\theta \nonumber \\
&=&\tilde{F}_{\Phi}(\phi_n)-\phi_n+I_n,
\end{eqnarray}
where the subscripts show the dependence on the parameters, the
subscript $\infty$ denotes $K \rightarrow \infty$, and the mean
angular coordinate difference depends on $A_n$ via
$\tilde{F}_{\Phi}(\phi_n)$. We define an ``instantaneous''
stochastic rotation number in the interval just before the
\textit{n}th to just before the (\textit{n} + 1)th impulses for the
phase equation as follows: 
\begin{eqnarray}
\Omega_{\infty,\epsilon,A_n,I_n}&=& \frac{1}{I_n}\int_0^1w_{\infty,A_n,I_n}(\phi_n)h_n(\phi_n)d\phi_n \nonumber \\
&=&1+\frac{1}{I_n} \int_0^1 (\tilde{F}_{\Phi}(\phi_n)-\phi_n) h_n(\phi_n) d\phi_n, \label{eq:StoRotInfInst}
\end{eqnarray}
where the subscripts of $\Omega_{\infty,\epsilon,A_n,I_n}$ represent
the dependence of the instantaneous stochastic rotation number on
the parameters $K \rightarrow \infty$, $\epsilon$, $A_n$, and $I_n$,
respectively. The instantaneous stochastic rotation number depends
on $\epsilon$ via $h_n(\phi_n)$. Note that 1:1 correspondence
between the set of input parameters $A_n$, $I_n$, and
$\Omega_{\infty,\epsilon,A_n,I_n}$ is achieved when $\epsilon$ is
fixed. If the impulse amplitude and interimpulse interval do not
vary with time, then the ``steady-state'' stochastic rotation number
is 
\begin{eqnarray}
\Omega_{\infty,\epsilon,A,I}&=&1+\frac{1}{I}\int_0^1 (\tilde{F}_{\Phi}(\phi)-\phi)h^*_{\infty,\epsilon,A,I}(\phi)d\phi, \label{eq:StoRotInfSteady}
\end{eqnarray}
where the subscripts $\infty$, $\epsilon$, $A$, and $I$ define $K
\rightarrow \infty$, the noise strength, a constant amplitude, and a
constant interimpulse interval, respectively;
$h^*_{\infty,\epsilon,A,I}$ is the invariant density of
$\mathbf{P}_{\infty,\epsilon,A,I}$.

Similarly, we define these stochastic rotation numbers for the full
equation. In this case, we have to include the difference caused by
the radial component of the density because the state points can
move across the entire phase plane. According to
Eq.~(8) in \cite{Yamanobe2011}, the state point just
before the $(n+1)$th impulse is distributed as
\begin{eqnarray}
n[(r,\theta);\mathbf{U}_{I_n}^{(0)}(r_n,\phi_n),\epsilon^2 \mathbf{\Sigma}_{I_n}(r_n,\phi_n)]+n[(-r,\theta+0.5);\mathbf{U}_{I_n}^{(0)}(r_n,\phi_n),\epsilon^2 \mathbf{\Sigma}_{I_n}(r_n,\phi_n)], \label{eq:KernelFullLifted}
\end{eqnarray}
where
$\mathbf{U}_{I_n}^{(0)}(r_n,\phi_n)=(R_{I_n}^{(0)}(r_n,\phi_n),\Theta_{I_n}^{(0)}(r_n,\phi_n))$
shows the explicit dependence on $(r_n,\phi_n)$ (the state point
just before the $n$th impulse), $r \in \{x;x>0,x\in \mathbf{R}\}$,
and $\theta \in \mathbf{R}$. Please note that the lifted angular coordinate is used in Eq. (\ref{eq:KernelFullLifted}).
The mean difference in the lifted
angular coordinates of the \textit{n}th and (\textit{n} + 1)th
impulses is 
\begin{eqnarray}
w_{K,\epsilon,A_n, I_n} (r_n,\phi_n) &=& \int_0^{\infty}\int_{-\infty}^{\infty}(\theta-\phi_n) n[(r,\theta);\mathbf{U}_{I_n}^{(0)}(r_n,\phi_n),\epsilon^2 \mathbf{\Sigma}_{I_n}(r_n,\phi_n)] d\theta dr \nonumber \\
&&+ \int_0^{\infty}\int_{-\infty}^{\infty}(\theta-\phi_n) n[(-r,\theta+0.5);\mathbf{U}_{I_n}^{(0)}(r_n,\phi_n),\epsilon^2 \mathbf{\Sigma}_{I_n}(r_n,\phi_n)]d\theta dr \nonumber \\
&=&(F_{\Phi}(r_n,\phi_n)-\phi_n+I_n)\int_0^{\infty} n[r;R_{I_n}^{(0)}(r_n,\phi_n),\epsilon^2 \mathbf{E}[A_{1RI_n}^2]]dr \nonumber \\
&&+ (F_{\Phi}(r_n,\phi_n)-(\phi_n+0.5)+I_n)\int_0^{\infty} n[-r;R_{I_n}^{(0)}(r_n,\phi_n),\epsilon^2 \mathbf{E}[A_{1RI_n}^2]]dr,
\end{eqnarray}
where the subscript of $w$ shows the dependence on the parameters $K$,
$\epsilon$, the $n$th impulse amplitude $A_n$, and the $n$th impulse
interimpulse interval $I_n$. The stochastic rotation number between
the \textit{n}th and (\textit{n} + 1)th impulses is 
\begin{eqnarray}
\Omega_{K,\epsilon,A_n,I_n}=\frac{1}{I_n}\int_0^1 \int_0^{\infty}
w_{K,\epsilon,A_n,I_n}(r_n,\phi_n) h_n(r_n,\phi_n) dr_n d\phi_n.
\label{eq:StoRot2nd}
\end{eqnarray}
This value is an instantaneous stochastic rotation number for the
full equation. We can also define the steady-state stochastic
rotation number for the full equation as follows: 
\begin{eqnarray}
\Omega_{K,\epsilon,A,I}=\frac{1}{I}\int_0^1 \int_0^{\infty} w_{K,\epsilon,A,I}(r,\phi) h_{K,\epsilon,A,I}^*(r,\phi) dr d\phi, \label{eq:StoRot1st}
\end{eqnarray}
where $h_{K,\epsilon,A,I}^*$ is the invariant density of
$\mathbf{P}_{K,\epsilon,A,I}$. Since $\Omega_{K,\epsilon,A_n,I_n}$
is the stochastic rotation number between the $n$th and $(n+1)$th
impulses, it corresponds to an impulse amplitude $A_n$ and input
interval $I_n$. Using this definition of the stochastic rotation
number, an instantaneous input rate corresponds to a specific output
rate. In this way, one can construct the instantaneous firing rate
of the Poincar\'{e} oscillator as a function of the instantaneous
input rate.

Furthermore, using the decomposition of $h_n(r_n,\phi_n)$ by
Eq.~(\ref{eq:DecompProduct}), Eq.~(\ref{eq:StoRot2nd}) can be
written as 
\begin{eqnarray}
\Omega_{K,\epsilon,A_n,I_n}&=&\int_0^1 \int_0^\infty w_{K,\epsilon,A_n,I_n}(r_n,\phi_n)h_{K,\epsilon,A_{n-1},I_{n-1}}^*(r_n,\phi_n)dr_n d\phi_n \nonumber \\
&+&\int_0^1 \int_0^\infty w_{K,\epsilon,A_n,I_n}(r_n,\phi_n) \sum_{i=0}^{n-2}(\prod_{j=0}^{n-i-2} Q_{K,\epsilon,A_{n-1-j},I_{n-1-j}})h_{K,\epsilon,A_i,I_i}^*(r_n,\phi_n)dr_n d\phi_n \nonumber\\
&+& \int_0^1 \int_0^\infty w_{K,\epsilon,A_n,I_n}(r_n,\phi_n)\prod_{l=0}^{n-2} Q_{K,\epsilon,A_{n-1-l},I_{n-1-l}}h_1(r_n,\phi_n)dr_n d\phi_n, \label{eq:FullStoRotDecomp}
\end{eqnarray}
where the operator $Q_{K,\epsilon,A,I}$ is from the original SPTO,
i.e., not from the discretized SPTO. Equation
(\ref{eq:FullStoRotDecomp}) tells us that the stochastic rotation
number is determined by three terms. The first term is a
contribution by the invariant density at the $(n-1)$th impulse. The
second term is the contribution produced by the invariant densities
corresponding to the input parameters
$\{A_1,I_1\},\ldots,\{A_{n-2},I_{n-2}\}$ with a weight determined by
the corresponding transient components of the SPTO. The third
component is from the initial density. The contribution of the
initial density is also weighted by the transient components of the
SPTO corresponding to given input impulses. Similarly, the
decomposition of Eq.~(\ref{eq:StoRotInfInst}) becomes
\begin{eqnarray}
\Omega_{\infty,\epsilon,A_n,I_n}&=&\int_0^1 \int_0^\infty w_{\infty,A_n,I_n}(\phi_n)h_{\infty,\epsilon,A_{n-1},I_{n-1}}^*(\phi_n)dr_n d\phi_n \nonumber \\
&+&\int_0^1 \int_0^\infty w_{\infty,A_n,I_n}(\phi_n) \sum_{i=0}^{n-2}(\prod_{j=0}^{n-i-2} Q_{\infty,\epsilon,A_{n-1-j},I_{n-1-j}})h_{\infty,\epsilon,A_i,I_i}^*(\phi_n)d\phi_n \nonumber\\
&+& \int_0^1 \int_0^\infty w_{\infty,A_n,I_n}(\phi_n)\prod_{l=0}^{n-2} Q_{\infty,\epsilon,A_{n-1-l},I_{n-1-l}}h_1(\phi_n) d\phi_n. \label{eq:ReducedStoRotDecomp}
\end{eqnarray}
In what follows, we use $\Omega_{K,\epsilon,A,I}$ for $K \in
(0,\infty]$, and we calculate $\Omega_{K,\epsilon,A,I}$ using the
same numerical integration methods explained in the calculation of
the spectral properties of the SPTO.

\subsection{Interspike interval density}
The ISI density is used to characterize the spiking activity of
neurons in the steady state. Nesse et al.
\cite{NesseClarkBressloff2007} calculated the ISI density of the
phase model with multiplicative noise by considering a population of
neuronal oscillators. We extend their idea to the case in which the
neuronal dynamics written in terms of the stochastic differential
equations. Specifically, we derive the ISI density for the reduced
model Eq.~(\ref{eq:RicPhase}). According to
\cite{NesseClarkBressloff2007}, the ISI density is derived in two
steps: 1) calculation of the relative spike density that gives the
time of the next input impulse arrival after a spike of the reduced
model, and 2) calculation of the conditional ISI density relative to
the first input impulse time. To calculate the ISI density, we make
the same two assumptions as Nesse et al.
\cite{NesseClarkBressloff2007}. The first assumption is that an
impulse does not produce the normalized angular coordinate shift
across unity. This is satisfied by assuming that an impulse makes an  
instantaneous horizontal shift by an amount equal to the impulse
amplitude. The second assumption is that the interimpulse interval
is large enough such that the normalized angular coordinate of the
next impulse is not behind that of the previous impulse.

We consider impulses with a constant amplitude and interimpulse
interval. For the calculation of the relative spike density, we set
the density just before the first input impulse to the invariant
density $h_{K,\epsilon,A,I}^*$, where $A$ and $I$ are the constant
input amplitude and interimpulse interval, respectively. We write
the relative spike density as $p_{Sp,Im}(\tau)$, where $\tau$ is the
relative time from the spike of the reduced model (Sp) to the next
impulse (Im). To calculate $p_{Sp,Im}(\tau)$, we have to take into
account the possibility that the phase model does not fire during
some input impulses (refer to Figure 1 in
\cite{NesseClarkBressloff2007}). Using the derivation method
outlined in \cite{NesseClarkBressloff2007}, $p_{Sp,Im}(\tau)$ for
the phase equation is given as 
\begin{eqnarray}
p_{Sp,Im}(\tau)&=&\sum_{j=1}^{\infty}p_{Sp,Im}^j(\tau), \tau \in [0,min\{I,1\}), \label{eq:PSpIm}
\end{eqnarray}
where
\begin{eqnarray}
\mathbf{P}_{K,\epsilon,A,I,p} h(\phi) &=& \int_0^1 n[\phi+p;\tilde{\Theta}_{I}^{(0)} (\phi),\epsilon^2 \Sigma_{\tilde{\Theta}_{I}}(\phi)] h(\phi) d\phi \\
\mathbf{P}_{K,\epsilon,A,I} &=& \sum_{p=-\infty}^{p=+\infty} \mathbf{P}_{K,\epsilon,A,I,p} \\
p_{Sp,Im}^k(\tau) &=& \mathbf{P}_{K,\epsilon,A,I,1}\mathbf{P}_{K,\epsilon,A,I,0}^{(k-1)}h_{K,\epsilon,A,I}^*(\tau), \label{eq:RelativeSpikeDens}
\end{eqnarray}
and $\mathbf{P}_{K,\epsilon,A,I,p}h(\phi)$ is the probability
density of neurons that fire $p$ times between two input impulses
with constant $A$ and $I$. The superscript $(k-1)$ in
Eq.~(\ref{eq:RelativeSpikeDens}) indicates that
$\mathbf{P}_{K,\epsilon,A,I,0}$ is raised to the $(k-1)$ power, and
$p_{Sp,Im}^j(\tau)$ is the density of the relative time from the
current spike to the next impulse after the preceding $j$ impulses.

The conditional ISI density relative to the first impulse time
$\tau$ is denoted by $p_{Sp,Sp}(T|\tau)$, where $T$ is the time
between successive output spikes of the reduced model relative to
$\tau$. When the reduced model receives a single impulse between
consecutive spikes, i.e., $T \in (\tau, I + \tau)$, we obtain the
following relations (refer to Figure 2 in
\cite{NesseClarkBressloff2007}): 
\begin{eqnarray}
T&=&I+\tau-\psi \label{eq:ReducedInterval}\\
1+\psi &=& \tilde{F}_{\Phi}(\tau) + I + \epsilon A_{1\tilde{\Theta}I}, \label{eq:ReducedPhase}
\end{eqnarray}
where we have approximated Eq.~(\ref{eq:RicPhase}) as 
\begin{eqnarray}
\tilde{\Theta}_I^{(\epsilon)} \sim \tilde{\Theta}_I^{(0)} + \epsilon A_{1\tilde{\Theta}I} = \tilde{F}_{\Phi}(\tau)+I+\epsilon A_{1\tilde{\Theta}I} \label{eq:ApproxReducedModel},
\end{eqnarray}
to derive Eq.~(\ref{eq:ReducedPhase}). Equation
(\ref{eq:ReducedInterval}) is a result of the relationship between
the ISI and impulses' normalized angular coordinate, while Eq.~(\ref{eq:ReducedPhase}) comes
from the relationship between the normalized angular coordinates $\psi$ and $\tau$. If the
model is given by stochastic differential equations, then the
approximation in Eq.~(\ref{eq:ApproxReducedModel}) is necessary so
that the density of $A_{1\tilde{\Theta}I}$ is included in the
following calculation. Using the probability density of
$A_{1\tilde{\Theta}I}$, i.e., the first term in Eq. (\ref{eq:DensityPhase}), we obtain the conditional ISI density for $T
\in (\tau, I + \tau)$: 
\begin{eqnarray}
p_{Sp,Sp}^1(T|\tau)&=&n[T;1+\tau-\tilde{F}_{\Phi}(\tau),\epsilon^2 \Sigma_{\tilde{\Theta}I}(\tau)].
\end{eqnarray}
In a similar manner, if two impulses exist between consecutive
spikes (i.e., $T\in [I+\tau,2I+\tau)$), then
\begin{eqnarray}
T &=& 2I + \tau - \psi \\
1+\psi &=& \tilde{F}_{\Phi}(\phi_1) + I + \epsilon A_{1\tilde{\Theta}I} \\
\phi_1 &=& \tilde{F}_{\Phi}(\tau) + I + \epsilon
A_{1\tilde{\Theta}I}.
\end{eqnarray}
The conditional density for this case is
\begin{eqnarray}
p_{Sp,Sp}^2(T|\tau) &=& \int_{I}^{\infty} n[T;t'+\tau-(\tilde{F}_{\Phi}(\phi_{1}(t'+\tau))-\phi_{1}(t'+\tau)),\epsilon^2\Sigma_{\tilde{\Theta}I}(\phi_{1}(t'+\tau))] \nonumber \\
&\times& p_{Sp,Sp}^{1}(t'+\tau|\tau)dt',
\end{eqnarray}
where $T\in [I+\tau,2I+\tau)$, and $\phi_1(t) = 1 - (t-\tau)+I$. In
general, the conditional ISI density for $T \in
[(j-1)I+\tau,jI+\tau)$ is 
\begin{eqnarray}
p_{Sp,Sp}^j(T|\tau) &=& \int_{(j-1)I}^{\infty} n[T;t'+\tau-(\tilde{F}_{\Phi}(\phi_{(j-1)}(t'+\tau))-\phi_{(j-1)}(t'+\tau)),\epsilon^2\Sigma_{\tilde{\Theta}I}(\phi_{(j-1)}(t'+\tau))] \nonumber \\
&\times& p_{Sp,Sp}^{(j-1)}(t'+\tau|\tau)dt',
\end{eqnarray}
where $\phi_j(t)=1 - (t-\tau) + jI$.

The conditional ISI density relative to the first impulse time
$\tau$ is then given by
\begin{eqnarray}
p_{Sp,Sp}(T|\tau)=p_{Sp,Sp}^j(T|\tau), \mspace{25mu}T \in [(j-1)I+\tau,jI+\tau).
\end{eqnarray}

For the case in which there are two consecutive spikes of the
reduced model, $\int_0^1 p_{Sp,Im}(T|\tau)d\tau$ reveals the
fraction of reduced models that receive at least one impulse, while
$1-\int_0^1 p_{Sp,Im}(\tau) d\tau$ reveals the fraction that receive
no impulse. Reduced models that receive no impulse generates spikes
with a mean period of $1$. If the reduced model receives two
impulses with zero amplitude at normalized angular coordinates of
$0$ and $1$, then Eq.~(\ref{eq:ApproxReducedModel}) means that the
following relation should hold: 
\begin{eqnarray}
1 = I + \epsilon A_{1\tilde{\Theta}I}.
\end{eqnarray}
Since these two zero-amplitude impulses are added at the normalized
angular coordinates of $0$ and $1$, the interimpulse interval $I$ is
equal to the ISI, $T=I$. Thus, the ISI density without impulses is
\begin{eqnarray}
n[T,1,\epsilon^2\Sigma_{\tilde{\Theta}I}(0)].
\end{eqnarray}
Note that we have not made an approximation with the
$\delta$-function as in \cite{NesseClarkBressloff2007}, since the
ISI density without impulses cannot be derived in this way. Thus,
the ISI density $p_{Sp,Sp}(T)$ becomes 
\begin{eqnarray}
p_{Sp,Sp}(T) = \int_0^1 p_{Sp,Sp}(T|\tau)p_{Sp,Im}(\tau)d\tau + n[T;1,\epsilon^2\Sigma_{\tilde{\Theta}I}(0)]\Bigl(1-\int_0^1 p_{Sp,Im}(\tau)d\tau \Bigr). \label{eq:ISIdensity}
\end{eqnarray}

For the full equation (Eq.~(\ref{eq:FullRicPolar})), the
variance-covariance matrix that corresponds to
$\Sigma_{\tilde{\Theta}I}(0)$ in Eq.~(\ref{eq:ISIdensity}) depends
on the radial variable. Thus, the ISI density cannot be derived as
in the case of the reduced model without further approximation. We
used the trapezoidal rule to calculate Eq.~(\ref{eq:ISIdensity}) for
the results presented here.

\subsection{Input impulses}
To examine the relation between the changing speed of input rate and 
the dynamics of the stochastic Poincar\'{e} oscillator, we
use impulses whose amplitudes are constant, and the instantaneous
input rate changes according to
\begin{eqnarray}
f_n &=& 1/I_n = f_{start} + \frac{(f_{end} - f_{start})}{N} (n-1) \label{eq:TimeVaryingImpulses} \\
f_{step} &=& (f_{end}-f_{start})/N, \label{eq:TimeVaryingImpulsesStep}
\end{eqnarray}
where $n=1,\cdots,N+1$. The input rate of the $1$st impulse and that
of the (\textit{N} + 1)th impulse are $f_{start}$ and $f_{end}$,
respectively. Here, \textit{N} determines the number of interimpulse
intervals, and $f_{step}$ denotes the step size of the input rate
change. By changing $f_{start}$, $f_{end}$, and \textit{N}, we can
investigate the response of the stochastic Poincar\'{e} oscillator
to time-varying impulses.

\section{Results}
\subsection{Stochastic kernels}
\begin{figure}[htbp]
\begin{center}
\includegraphics[scale=1]{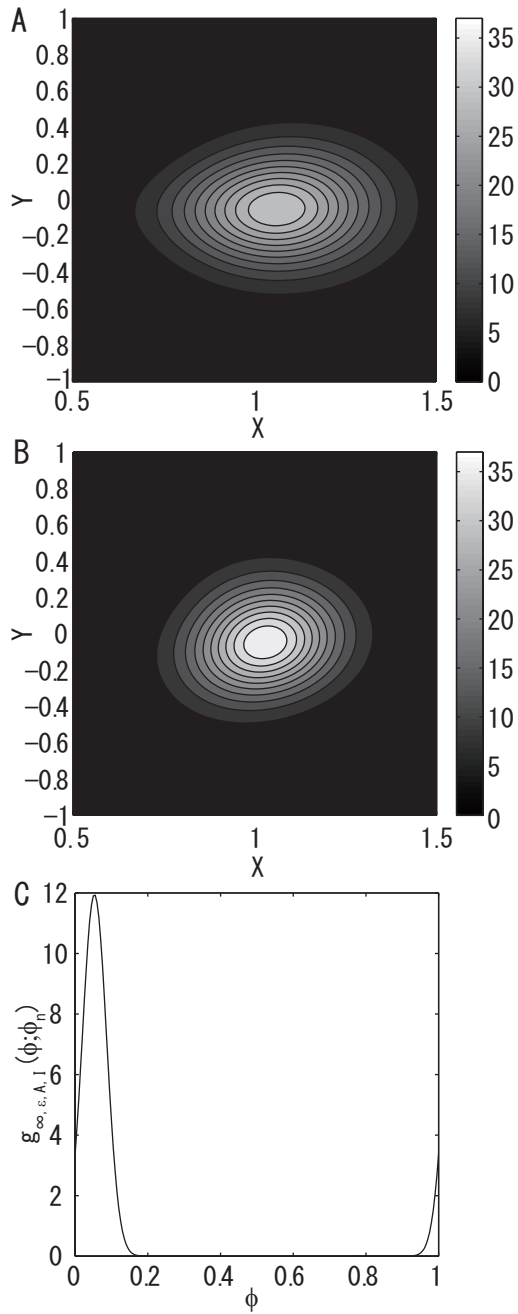}
\end{center}
\caption{ {\bf Stochastic kernels.} Stochastic kernels with (A)
$K=0.25$, (B) $K=1$, and (C) $K=\infty$. (A) and (B) were calculated
using Eq.~(8) in \cite{Yamanobe2011} and (C) was calculated using
Eq.~(\ref{eq:ReducedTranProbDens}). The parameters for the
stochastic kernels were $A=0.95$, $I=0.95$, $\epsilon=0.3$, with
initial conditions of $(r_1,\phi_1)=(0.3,0.2)$ for (A) and (B) and
$\phi_1=0.2$ for (C). In (A) and (B), the abscissa and ordinate are
the membrane potential and refractoriness of the stochastic
Poincar\'{e} oscillator, respectively. The color bar shows the
probability density of the stochastic kernel. In (C), the abscissa
is the normalized angular coordinate, and the ordinate is the
probability density of the stochastic kernel.} \label{Figure_label2}
\end{figure}

Figure \ref{Figure_label2} shows the stochastic kernels for
$K=0.25$, $1$, and $\infty$ calculated from the same initial state
(in the case of $K=\infty$, the same initial normalized angular
coordinate as in the finite $K$ case is used). Comparing the
stochastic kernels for the finite relaxation rates, $K=0.25$ and
$1$, we see that the shape of the kernel with the larger relaxation
rate has induced a sharper unimodal density.

\subsection{Density evolution}
\begin{figure}[htbp]
\begin{center}
\includegraphics[scale=1]{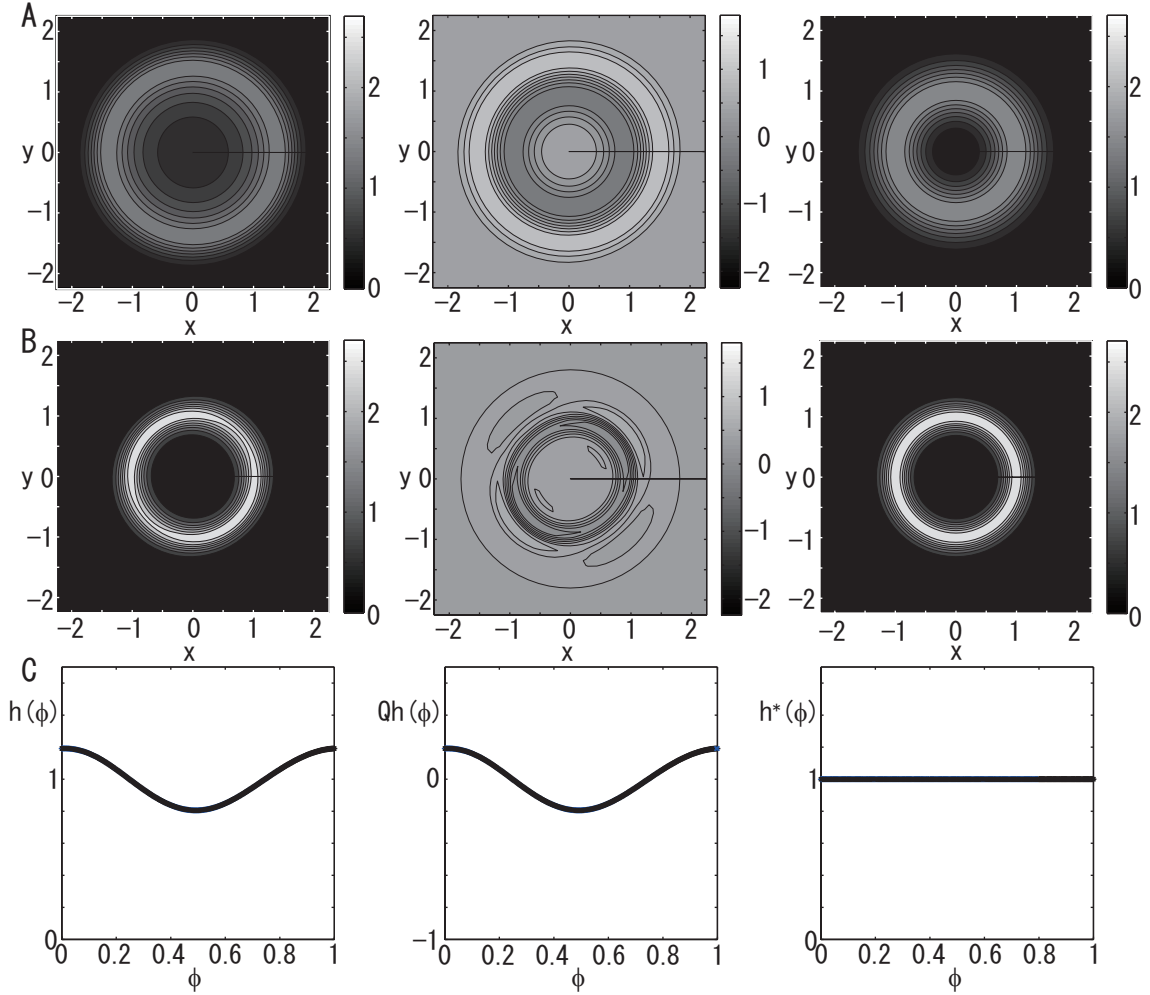}
\end{center}
\caption{ {\bf Statistical global dynamics of the stochastic
Poincar\'{e} oscillator without any impulse.} The statistical phase
plane dynamics of the stochastic Poincar\'{e} oscillator are shown
by the probability density (left column; calculated using
Eq.~(\ref{eq:DiscretizedFullMarkovOperator}) for (A) and (B) and
Eq.~(\ref{eq:ReducedMarkovOperatorTrapz}) for (C)), the
corresponding transient components (middle column;
Eq.~(\ref{eq:SpecDecomp})), and the corresponding invariant density
(right column; Eq.~(\ref{eq:SpecDecomp})) without any impulses,
i.e., $A=0$ after a time interval of $I=2.75$ for (A) and (B) and
$I=2.75$ for (C). The relaxation rates are (A) $K=0.25$, (B) $K=1$,
and (C) $K=\infty$. For (A) and (B), the initial density
$h_1(r_1,\phi_1)$ was a uniform distribution with a support of
$(r_1,\phi_1) \in (0,3.5]\times[0,1)$, and for (C), the initial
density $h_1(\phi_1)$ was a uniform distribution with a support of
$\phi_1 \in (0.2,0.8]$. In all cases, $\epsilon=0.3$. The color bar shows the
probability density or the corresponding transient component. In (A) and
(B), abscissae and ordinates are the membrane potential and
refractoriness of the stochastic Poincar\'{e} oscillator,
respectively. In (C), the abscissa is the normalized angular
coordinate and the ordinate is probability density (left and right panels) or
transient component (middle panel).} \label{Figure_label3}
\end{figure}

If $A=0$, then the SPTO describes a density evolution that reflects
the dynamics of the stochastic Poincar\'{e} oscillator itself. The
spectral decomposition of the SPTO also decomposes the density into
a transient component and an invariant density. Figure
\ref{Figure_label3} shows the densities and corresponding transient
components and invariant densities for different values of
\textit{K}. (The evolutions of the densities and transient
components are shown in Video S1 \cite{VideoS1}.) Since the deterministic
Poincar\'{e} oscillator has a stable limit cycle, the densities
evolve toward this limit cycle and then converge to the
corresponding invariant densities that are distributed around the
limit cycle. The convergence speed depends on the relaxation rate of
the stochastic Poincar\'{e} oscillator. That is, a larger relaxation
rate is associated with a smaller transient component.

\subsection{Spectral properties of the SPTO}

If $K \rightarrow \infty$ and $\epsilon=0$, the Poincar\'{e}
oscillator is known to exhibit typical structure in response to
impulses with a constant \textit{A} and \textit{I}. In particular,
for $|A|<1$, the dynamics of the stochastic Poincar\'{e} oscillator
are described by the 1D phase transition curve of
Eq.~(\ref{eq:1DPTC}), which is an invertible diffeomorphism of the
circle. For this case, the responses are classified into two
categories: a phase locking, in which \textit{q} impulses correspond
to \textit{p} spikes (\textit{p} and \textit{q} are integer values),
and a quasi-periodic response, where one impulse rotates the
Poincar\'{e} oscillator an irrational number of times
\cite{GlassSun1994}.

\begin{figure}[htbp]
\begin{center}
\includegraphics[scale=1]{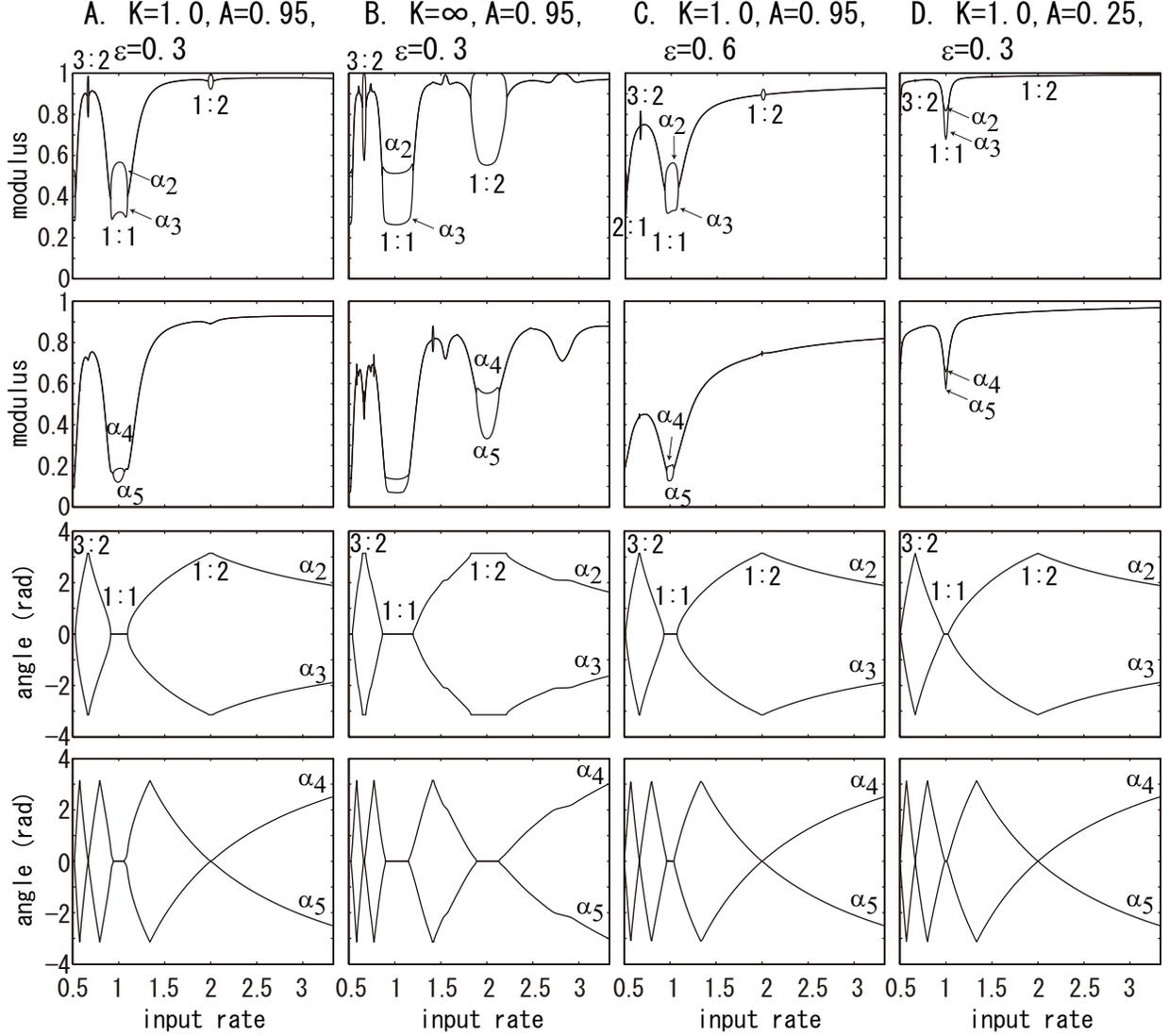}
\end{center}
\caption{ {\bf Moduli and angles of the eigenvalues of the
discretized SPTO as a function of the input rate.} The eigenvalues
were calculated using Eq.~(\ref{eq:DiscretizedFullMarkovOperator})
for finite $K$ and Eq.~(\ref{eq:ReducedMarkovOperatorTrapz}) for
infinite $K$. For each set of parameters, the moduli and angles of
the second to fifth eigenvalues are plotted. Some stochastic
phase-locking regions are labeled with their locking ratio. The
parameters are shown in each panel. The input rate is on the
abscissa and the moduli or angles of the eigenvalues of the
discretized SPTO are on the ordinate.} \label{Figure_label4}
\end{figure}

Using the definition of stochastic phase locking based on the
eigenvalues of the discretized SPTO, we evaluated the effect of the
relaxation rate, noise strength, impulse amplitude, and inverse of
the interimpulse interval (input rate) on the response of the
impulse-driven stochastic Poincar\'{e} oscillator. Figure
\ref{Figure_label4}A and B show the moduli and angles of the
eigenvalues as a function of the input rate for different values of
\textit{K}. The overall trend of the moduli for both cases was to
increase as the input rate increased, and some larger stochastic
phase-locking regions survived even in the presence of noise. As
\textit{K} increased, there was an overall decrease in the moduli of
the eigenvalues, and the stochastic phase-locking regions became
wider. In some stochastic phase-locking regions, the modulus of the
second eigenvalue became larger as $K$ increased (for example, in
the 1:2 and 3:2 stochastic phase-locking regions). Furthermore, in
some regions, the modulus of the second eigenvalue took a value that
was slightly smaller than 1, as can be seen in the 1:2 stochastic
phase-locking region. This indicates that the corresponding
eigenfunction largely affects the dynamics in this region. In fact,
the density tends to rotate in the phase plane, and the response can
have a long transient regime even though this phenomenon depends on
the initial density.

A comparison of Figures \ref{Figure_label4}A and
\ref{Figure_label4}C reveals how the moduli and angles of the
eigenvalues of the discretized SPTO change in response to an
increase in the noise strength, i.e., there was an overall decrease
in the moduli of the eigenvalues but an increase in the modulus of the 
second eigenvalue around the 1:1 and 2:1 stochastic phase-locking regions. The
stochastic phase-locking regions also became narrower. Thus, the
detailed stochastic bifurcation structure disappears when the noise
strength is increased.

Furthermore, the dependence of the moduli and angles of the
eigenvalues of the discretized SPTO on the impulse amplitude can be
seen in a comparison of Figures \ref{Figure_label4}A and
\ref{Figure_label4}D. In this study, we concentrated on the impulse
amplitude $|A| < 1$, since an Arnold tongue structure exists in this
range, at least for infinite $K$ \cite{GlassSun1994}, and this
structure is a general structure of nonlinear oscillators. A
comparison shows that the stochastic phase-locking region narrows
when the impulse amplitude decreases. This is similar to the
narrowing of the deterministic phase-locking region that is seen
when the amplitude of the impulse decreases for $|A|<1$ and infinite $K$. An increase
in the impulse amplitude resulted in an overall decrease in the
moduli of the eigenvalues. However, the modulus of the second eigenvalue 
of the 1:2 and 3:2 stochastic phase-locking regions increased as an increase in 
the impulse amplitude.

\begin{figure}[htbp]
\begin{center}
\includegraphics[scale=1]{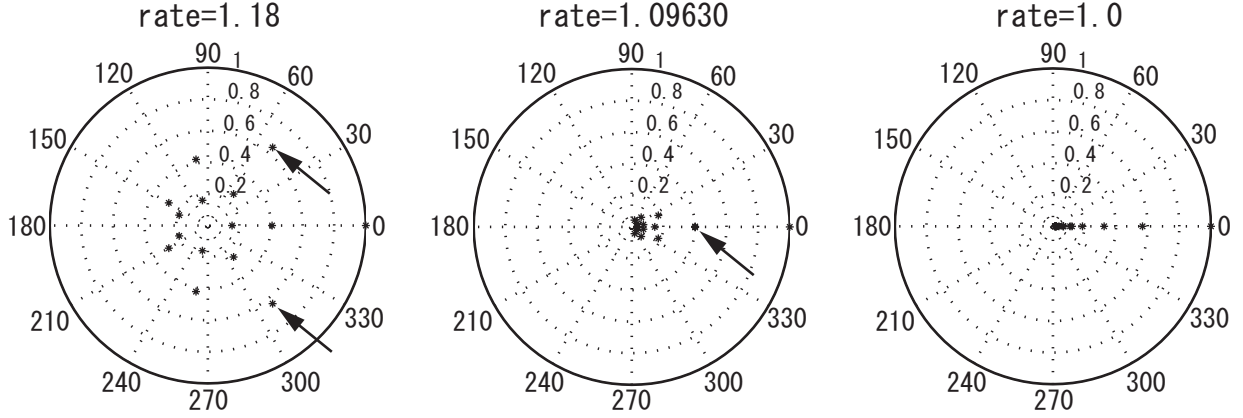}
\end{center}
\caption{ {\bf Stochastic saddle-node bifurcation.} The first 15
eigenvalues of the discretized SPTO
(Eq.~(\ref{eq:DiscretizedFullMarkovOperator})) are plotted in the
complex plane for the input rate indicated in each panel. The other
parameters were $A=0.95$, $\epsilon=0.3$, and $K=1$. In the
left-hand panel, the arrows indicate the second and third
eigenvalues. In the middle panel, the arrow indicates the second
eigenvalue. As the input rate decreased, the second and third
eigenvalues coincide, which is the stochastic saddle-node
bifurcation. For more details about the stochastic bifurcation, see
the text. } \label{Figure_label5}
\end{figure}

To understand the type of stochastic bifurcation of the stochastic
Poincar\'{e} oscillator with a finite relaxation rate, Figure
\ref{Figure_label5} shows the distribution of the first 15
eigenvalues of the discretized SPTO for finite $K$ ($K=1$). Some
eigenvalues, whose moduli were not in the vicinity of zero and less
than 1, are located on the positive part of the $x$-axis throughout
the stochastic bifurcation. We did not observe these eigenvalues in
the case of the SPTO for $K=\infty$. In relation to the definition
of the stochastic bifurcation in terms of the distribution of the
eigenvalues of the transition operator (in our case, this
corresponds to the SPTO) \cite{BorisyukRassoulAgha}, we also
examined the eigenvalue distribution as a function of the input
rate. We found that as the input rate decreased, the eigenvalue
distribution exhibited ``zipping'' behavior, as shown in Figure
\ref{Figure_label5}. That is, in a similar manner to closing a zip,
the complex eigenvalues converge to real values as the input rate
increases (left to right panel). It would seem that several spirals
exist around the zero point but we were unable to determine the
detailed structure around the zero point because the accuracy of the
eigenvalues in this region was insufficient.

In the neighborhood of the stochastic 1:1 phase-locking region, the
second and third eigenvalues were complex conjugates that correspond
to stochastic quasi-periodic responses (these eigenvalues are
indicated by the arrows in Figure \ref{Figure_label5} left panel).
For an input rate of 1.09630, the second eigenvalue was real, and
for an input rate of 1.0, eigenvalues with moduli smaller than that
of the second eigenvalue were real. Thus, a stochastic bifurcation
occurs as the input rate decreases. We checked and confirmed that
similar changes occur with a change in the input rate in the 1:2
stochastic phase-locking case. Thus, the Poincar\'{e} oscillator
shows a stochastic saddle-node bifurcation, as defined in
\cite{DoiInoueKumagai1998}. In addition, the shapes of the invariant
densities did not change abruptly as a function of the input rate as
has been reported for other systems
\cite{TatenoDoiSatoRicciardi1995,DoiInoueKumagai1998,Tateno1998,TatenoJimbo2000,Tateno2002}
(the change in the invariant density as a function of the input rate
is shown in Video S2 \cite{VideoS2}).

\subsection{Stochastic rotation number}

The firing rate is an important statistic in spike-train analysis,
and here, the counterparts of the firing rate are the instantaneous
and steady-state stochastic rotation numbers. We used these numbers
to understand the difference between the responses in the steady and
transient states.

The curves in Figure \ref{Figure_label6} show how the
steady-state stochastic rotation number changes as a function of the
input rate. In a \textit{p}:\textit{q} stochastic phase-locking
region, the slope of the stochastic rotation number was close to
\textit{p}/\textit{q}. Since the invariant density changed smoothly
as a function of the input rate, the steady-state stochastic
rotation number did not show any abrupt changes at the edges of the
stochastic phase-locking regions. As \textit{K} decreased, the
stochastic phase-locking regions narrower, and regions in which the
slope was close to \textit{p}/\textit{q} also narrowed (compare the
 curve in Figure \ref{Figure_label6}A for $K=1$ with that in
\ref{Figure_label6}F for $K=\infty$). Furthermore, an increase in
the noise strength and a decrease in the impulse amplitude in the
$|A|<1$ range narrowed regions in which the slope was close to $p/q$
(compare Figure \ref{Figure_label6}A with \ref{Figure_label6}H for
the noise strength and Figure \ref{Figure_label6}A with
\ref{Figure_label6}I for the impulse amplitude).

If the time-varying impulses defined by
Eq.~(\ref{eq:TimeVaryingImpulses}) are added, then the response will
have different properties from those of the steady state. We set
$f_{step}$ to different values by changing \textit{N}, $f_{start}$,
and $f_{end}$ in Eq.~(\ref{eq:TimeVaryingImpulsesStep}) to
investigate the dependence of the response on $f_{step}$. Figure
\ref{Figure_label6}A and B show the instantaneous stochastic
rotation number (asterisks) for different signs of $f_{step}$, but
the absolute values of $f_{step}$, as well as the minimum and
maximum input rates, were the same in both cases. Depending on
whether the input rate increased or decreased, the stochastic
rotation number showed different behavior. For example, if the input
rate entered the 1:1 stochastic phase-locking region, then the
instantaneous stochastic rotation number crossed the curve,
that is, the steady-state stochastic rotation number, and took a
smaller (larger) value than the steady-state stochastic rotation
number as the input rate increased (decreased).

Figure \ref{Figure_label6}C shows the instantaneous stochastic
rotation numbers that originate from the different initial densities
from Figure \ref{Figure_label6}A for the impulses whose
input rates start from $0.5$. After the first five impulses, the
effect of the initial density was negligible.

If $|f_{step}|$ was small enough, then the instantaneous stochastic
rotation number took a similar value to that of the steady state
even though the modulus of the second eigenvalue was slightly
smaller than 1; as seen, for example, in the 1:2 stochastic
phase-locking region (Figure \ref{Figure_label6}E). Furthermore, since the product of the discretized
SPTOs is weakly ergodic, the initial density is ``forgotten'' if the
number of impulses is large enough, and this leads to the third term
in Eq.~(\ref{eq:FullStoRotDecomp}) going to zero. Moreover, the
invariant density changed smoothly as a function of the input rate,
and thus the density $h_{K,\epsilon,A_{n-1},I_{n-1}}^*$ was similar
to $h_{K,\epsilon,A_n,I_n}^*$ in Eq.~(\ref{eq:FullStoRotDecomp}). 
Furthermore, according to Eq.~(\ref{eq:SecondTermProduct}), 
the individual terms in the second term of Eq.~(\ref{eq:FullStoRotDecomp}) 
become small if $|f_{step}|$
is small enough. This makes the instantaneous stochastic rotation number similar to
the steady-state counterpart.

In the region where the modulus of the second eigenvalue was
slightly less than 1, the response produced by the discretized SPTO
had a large transient component. This property induced a response
that was different from the steady-state response even though the
response depends on the initial density (Figure \ref{Figure_label6},
except for panel E). For example, in Figure \ref{Figure_label6} A,
B, and C, the instantaneous stochastic rotation number oscillated
and largely different from the steady-state counterpart when the
instantaneous input rate was roughly within $[1.5, 1/0.3]$. The
larger eigenvalues of the discretized SPTO in this range take mostly
complex values, and their absolute values were close to 1. This
property leads to a difference in the instantaneous and steady-state
stochastic rotation numbers and the oscillation of the instantaneous
stochastic rotation number.

If we fix the range of the input rate and $f_{step}$ and examine
results for a decrease in the impulse amplitude
and an increase in the noise strength, we see that this leads to a 
smaller variation in the instantaneous stochastic rotation number
(compare Figure \ref{Figure_label6}A with \ref{Figure_label6}I, 
and \ref{Figure_label6}H, respectively). We also investigated the 
dependence of the instantaneous stochastic rotation number 
by changing the relaxation rate (Figure \ref{Figure_label6}A for $K=1$, 
\ref{Figure_label6}F for $K=\infty$ and \ref{Figure_label6}G for $K=0.6$). 
As the relaxation rate increased beyond $0.4$, 
the variation of the instantaneous stochastic rotation number around the input rate 
$2$ decreased for finite $K$ (we observed this tendency by checking $K \in [0.1,1.2]$ with 
step $0.1$). 
However, the variation around the input rate $2$ increased 
as the relaxation rate increased if we compared Figure \ref{Figure_label6}A 
with Figure \ref{Figure_label6}F.

\subsection{Interspike interval density}
To understand the response in the steady state, we calculated the
ISI density. Figure \ref{Figure_label7} shows the ISI densities for
$K=\infty$ and various input rates. In the $1:1$ stochastic
phase-locking region, the ISI density was unimodal, with the mean
ISI similar to the input period. As the input rate decreased, the peak
shifted to larger ISI values and decreased in magnitude; a plateau
also appeared (Figure
\ref{Figure_label7}B). A further decrease in the input rate led to
the appearance of three local maxima in the ISI density (Figure
\ref{Figure_label7}C). Figure \ref{Figure_label7}B and C show the
densities outside the 1:1 stochastic phase-locking region, and there
is a stochastic bifurcation point between the densities shown in
Figure \ref{Figure_label7}A and \ref{Figure_label7}B. However, the
change in the ISI density was smooth because the ISI density is
based on the invariant density of the corresponding SPTO. The
accuracy of the calculated ISI density depended on the input rate,
and especially for input rates higher than $1$, calculating the ISI
was difficult within the scope of the reduced model; one reason
being that the convergence of the summation $\Sigma_{j=1}^{\infty}
p_{Sp,Im}(\tau)$ in Eq.~(\ref{eq:PSpIm}) was slow.

\subsection{Dependence of the current instantaneous stochastic rotation number on the past activity of the stochastic Poincar\'{e} oscillator}
To understand the effects of the past activity of the stochastic
Poincar\'{e} oscillator on the current instantaneous stochastic
rotation number, we calculated the past components that determine
the current instantaneous stochastic rotation number. As an example,
let us consider the instantaneous stochastic rotation numbers
corresponding to five impulses with a fixed impulse amplitude $A$
and four interimpulse intervals denoted by $I_1$, $I_2$, $I_3$, and
$I_4$. The equation that determines the instantaneous stochastic
rotation number at the last interimpulse interval is 
\begin{eqnarray}
\Omega_{K,\epsilon,A,I_4}&=&\int_0^1 \int_0^{\infty} w_{K,\epsilon,A,I_4}(r,\phi)h_{K,\epsilon,A,I_3}^*(r,\phi)dr d\phi \nonumber \\
&+&\int_0^1 \int_0^{\infty} w_{K,\epsilon,A,I_4}(r,\phi)Q_{K,\epsilon,A,I_3}h_{K,\epsilon,A,I_2}^*(r,\phi)dr d\phi \nonumber \\
&+&\int_0^1 \int_0^{\infty} w_{K,\epsilon,A,I_4}(r,\phi)Q_{K,\epsilon,A,I_3}Q_{K,\epsilon,A,I_2}h_{K,\epsilon,A,I_1}^*(r,\phi)dr d\phi \nonumber \\
&+&\int_0^1 \int_0^{\infty} w_{K,\epsilon,A,I_4}(r,\phi)Q_{K,\epsilon,A,I_3}Q_{K,\epsilon,A,I_2}Q_{K,\epsilon,A,I_1} h_1(r,\phi) dr d\phi. \label{eq:StoI4I3I2I1}
\end{eqnarray}
The first term in Eq.~(\ref{eq:StoI4I3I2I1}) is determined by the
invariant density that corresponds to the third impulse and the mean
lifted angular coordinate difference at the fourth impulse. Thus,
the first term depends on $I_3$ and $I_4$. We assign $I_3$, that is,
the oldest interimpulse interval in this term, to this term to show
that it depends on the past activity from the third impulse.
Similarly, the second term of Eq.~(\ref{eq:StoI4I3I2I1}) depends on
$I_2$, $I_3$, and $I_4$, and we assign $I_2$ to the second term. The
sum of the third and fourth terms of Eq.~(\ref{eq:StoI4I3I2I1})
depends on $I_1$, $I_2$, $I_3$, and $I_4$, and we assign $I_1$ to
the sum of these terms. This correspondence shows the dependence of
the current instantaneous stochastic rotation number on the past
activity of the stochastic Poincar\'{e} oscillator, and this can
also be seen in Figure \ref{Figure_label8} as a function of the
input rate.

The components of the instantaneous stochastic rotation number in
response to impulses with a fixed amplitude are shown in Figure
\ref{Figure_label8} (the components are denoted by asterisks, and
the filled square and square show the current instantaneous stochastic
rotation number). Figure \ref{Figure_label8}A and B use the same set
of input rates but different signs of $|f_{step}|$, i.e., the input
rates change in opposite directions. The difference in the sign of
$f_{step}$ induced a different dependence of the current
instantaneous stochastic rotation number on the past activity of the
stochastic Poincar\'{e} oscillator: Figure \ref{Figure_label8}A
shows a relatively larger component value for input rates between
$\sim$$1.0$ and $2.0$ and a weaker dependence on impulses near the
current input rate is seen, except for the component just before the
final input rate at which the current output rate is shown. In
contrast, Figure \ref{Figure_label8}B shows the instantaneous
stochastic rotation number depended only on components near the
current input rate.

As the number of the impulses increased, the variation in the
components tended to decrease (compare Figure \ref{Figure_label8}A
with \ref{Figure_label8}C and \ref{Figure_label8}D). Figure
\ref{Figure_label8}D shows the components of the instantaneous
stochastic rotation number corresponding to Figure
\ref{Figure_label6}E. As the number of impulses determined by
Eq.~(\ref{eq:TimeVaryingImpulses}) increases, the difference between
adjacent invariant densities corresponds to the input rate, and
components that depend on the initial density decrease in magnitude.
Thus, the contribution of components, apart from the component just
before the current input rate, subsequently decrease.

The instantaneous stochastic rotation number and components at the
5th impulse are shown in Figure \ref{Figure_label8}E for two
different initial densities (the filled square and asterisks
corresponds to the case in Figure \ref{Figure_label6}A and the 
square and asterisks to that in \ref{Figure_label6}C). For some input rates, the 
components for the two cases are superimposed. This
difference would appear to originate from the difference in the
initial densities.

Furthermore, we checked the dependence of the current instantaneous 
stochastic rotation number on the past activity by changing the relaxation rate 
(Figure \ref{Figure_label8}F and \ref{Figure_label8}G), the noise strength 
(Figure \ref{Figure_label8}H),  and the impulse amplitude  
(Figure \ref{Figure_label8}I and \ref{Figure_label8}J), respectively. 
For the finite relaxation rate, an increase of the relaxation rate led 
to the lower variation of the past components except the components 
corresponding to the 17th - 20th impulses (compare Figure \ref{Figure_label8}A 
with \ref{Figure_label8}G). However, for the infinite relaxation rate (compare Figure 
\ref{Figure_label8}A and \ref{Figure_label8}F), the variation of the 
components at from the 8th impulse to the 16th impulse was larger than 
that for the finite relaxation rate and that of the components corresponding 
to the 17th - 20th impulses decreased. An increase in the noise strength 
induced that, as can be seen in a comparison of Figure \ref{Figure_label8}A and 
\ref{Figure_label8}H, a larger noise strength induced a smaller variation in 
the components. As the impulse amplitude increased until about $0.45$ 
(compare Figure \ref{Figure_label8}I and Figure \ref{Figure_label8}J), 
the variability of the past components around the input rate $1$ increased. 
If the impulse amplitude increased beyond about $0.45$ (compare 
Figure \ref{Figure_label8}A and \ref{Figure_label8}J), the overall trend 
of the variability around the input rate $1$ decreased, and the variability 
of some components after the 9th impulse increased slightly. 

\section{Discussion}

The transient regime should be short enough to encode information 
in the spike pattern in nervous systems. The length of the 
transient regime becomes an indicator of the extent of the 
dependence on the past neural activity. 
By decomposing the current instantaneous
stochastic rotation number into the past activity components, we
were able to tackle this problem. The results showed that the
components of the current instantaneous stochastic rotation number
can be negative, and components far from the current input rate can
affect the current instantaneous stochastic rotation number. 
Some components depend on the difference between the invariant densities
of adjacent discretized SPTOs in the product of discretized SPTOs,
which determines the density in the equation of the current
instantaneous stochastic rotation number. A larger difference
between invariant densities may increase the values of the
corresponding components. Furthermore, the product of discretized
SPTOs shows weak ergodicity, that is, the stochastic Poincar\'{e}
oscillator can forget the initial density, and equally, the current
instantaneous stochastic rotation number can forget the initial density or initial
condition. This situation may arise in nervous systems if a neuron
receives a sufficient number of impulses. 

The kernel density estimation is a method to estimate spike rate \cite{Parzen1962,Rosenblatt1956,Sanderson1980,Richmond1990a,Nawrot1999,Shimazaki2010}.
In this method, spike train is convoluted with a kernel function to estimate spike 
rate. Further analysis of the past activity components 
of the current instantaneous stochastic rotation number might 
lead to an adequate selection of the kernel function, 
that is usually a nonnegative function, and its width 
to calculate spike rate.

\begin{figure}[htbp]
\begin{center}
\includegraphics[scale=1]{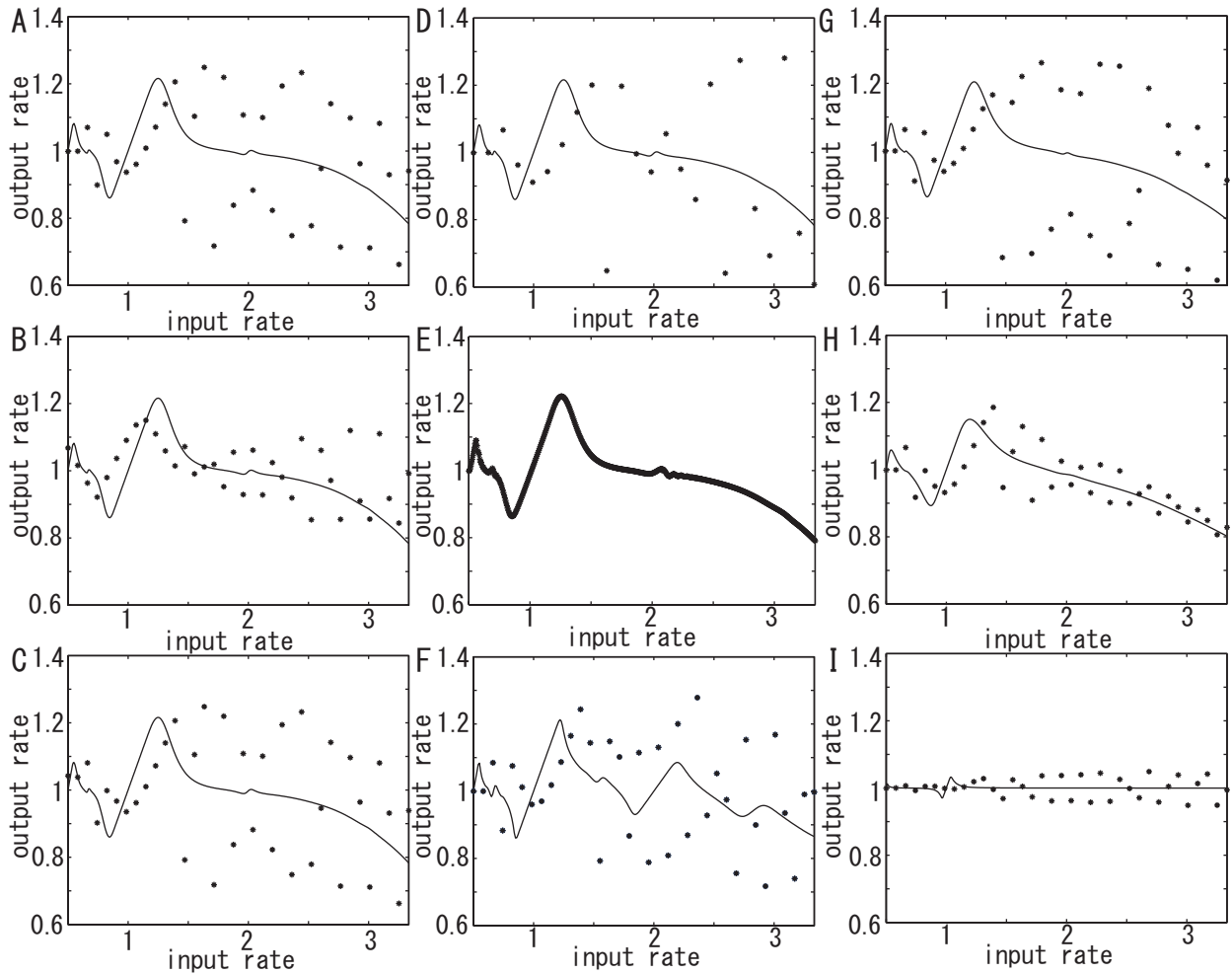}
\end{center}
\caption{ {\bf Stochastic rotation number as a function of the input
rate.} The curves show the steady-state stochastic rotation
number as a function of the input rate (calculated using
Eq.~(\ref{eq:StoRotInfSteady}) or (\ref{eq:StoRot1st})) and the asterisks 
show the instantaneous stochastic rotation number as a function
of the input rate for different parameters (calculated using
Eq.~(\ref{eq:StoRotInfInst}) or (\ref{eq:StoRot2nd})). We varied the
relaxation rate, noise strength, and amplitude of the input impulse.
(A) Standard input--output rate plot. The starting and final input
rates were $f_{start} = 0.5$ and $f_{end}=1/0.3$, $36$ impulses ($N=35$)
with an amplitude of $A=0.95$ were considered along with a
relaxation rate of $K=1$ and noise strength of $\epsilon=0.3$. The
initial density $h_1(r_1,\phi_1)$ was a uniform distribution with a
support of $(r_1,\phi_1) \in (0,3.5]\times[0,1)$. In (B)--(I), the
effects of various parameters on the stochastic rotation numbers are
calculated by varying one parameter while the other parameters and
initial density are the same as those in (A). Stochastic rotation
numbers for (B) $f_{start} = 1/0.3$ and $f_{end} = 0.5$ to
investigate the effect of $f_{step}$ in
Eq.~(\ref{eq:TimeVaryingImpulsesStep}), (C) the initial density with
a support of $(r_1,\phi_1) \in (0,3.5]\times[0.25,0.75)$ to examine
the effect of the initial density, (D) $N=23$, (E) $N=500$, (F)
$K=\infty$, (G) $K=0.6$, (H) $\epsilon = 0.6$, and (I) $A=0.25$.}
\label{Figure_label6}
\end{figure}

\begin{figure}[htbp]
\begin{center}
\includegraphics[scale=1]{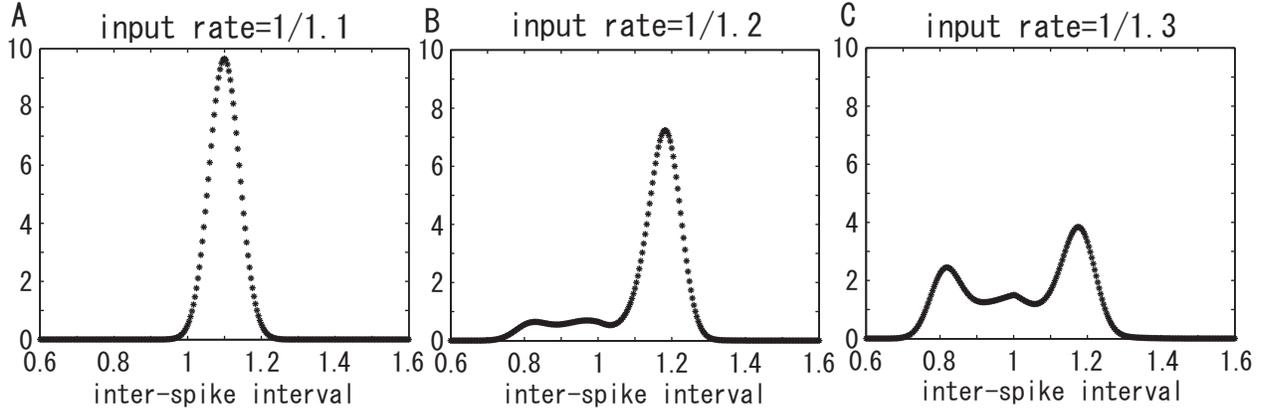}
\end{center}
\caption{ {\bf Interspike interval density.} Interspike interval
densities of the reduced model calculated using
Eq.~(\ref{eq:ISIdensity}). (A) The interspike interval density in
the $1:1$ stochastic phase locking region. (B) and (C) show the
densities outside the $1:1$ stochastic phase locking region for
$I=1.2$ and $I=1.3$, respectively. There is a stochastic bifurcation
point between (A) and (B), but the interspike interval density
changed smoothly because the invariant density of the corresponding
SPTO changed smoothly. Plots are shown for $K=\infty$, $A=0.95$, and
$\epsilon=0.3$. } \label{Figure_label7}
\end{figure}

As shown by an examination of the stochastic rotation number, the
steady-state and transient responses can be different. In a related
experiment it was found that the response of a pacemaker neuron in
crayfish was different depending on the past inputs
\cite{SegundoVibertStiberHanneton1995a,SegundoVibertStiberHanneton1995b,SegundoVibertStiber1998}.
We used the definition of the instantaneous stochastic rotation
number as given by Eqs.~(\ref{eq:StoRotInfInst}) and
(\ref{eq:StoRot2nd}), since we examined the input--output firing
rate relationship as shown in Figure \ref{Figure_label6}. This is a
natural extension of the stochastic rotation number in the steady
state to that in the transient state. This is also required in
experiments to establish the input--output rate relationship. In
fact, past studies typically relate the input and output rates
empirically (see
\cite{SegundoVibertStiberHanneton1995a,SegundoVibertStiberHanneton1995b}
for examples). To define the SPTO and instantaneous stochastic
rotation number for a continuous input over a continuous time period
leads to the firing rate over that continuous time period, and this
is a topic for future study. Analysis using the continuous-time
version of the SPTO will yield the continuous dependence of the
current firing rate on the past activity of the neuron model and
give some insight into information coding in nervous systems.

For the model dynamics, we showed that the contribution of the past
activity of the stochastic Poincar\'{e} oscillator to the current
density was defined by Eq.~(\ref{eq:DecompProduct}). The product of
the discretized SPTOs determines the response, i.e., the current
density, of the neuronal oscillator to time-varying impulses.
Alternatively, the effect of the terms in
Eq.~(\ref{eq:DecompProduct}) can be examined by using the 1-norm, if
necessary, and these terms show the dependence of the current
density on the past neuronal activity.

A shorter dependence on the past activity is necessary for carrying
information via a spike pattern, and a longer dependence on the past
activity might lead to information being carried with a weighted average
of the number of spikes that is determined by the transient dynamics of the
neuron. To understand the mechanism behind this dependence, it is
necessary to understand the spectral structure of the discretized
SPTO. A response of the stochastic Poincar\'{e} oscillator, or a
product of the discretized SPTOs, depends on the discretized SPTOs
selected by the input parameters. Hence, it is an important problem
to understand how each discretized SPTO in a product of discretized
SPTOs affects the spectral property of the product. Since the
discretized SPTO is a non-commutative matrix, the order of the
multiplication also affects this spectral property.

As mentioned in the introduction, we focused on the transient 
dynamics of a neuronal oscillator. Usually, conventional statistics
 such as the spike train power spectrum, Fano factor of the spike 
count, and ISI density assume the oscillator is in the steady state. 
These statistics are not adequate for quantifying the transient 
dynamics of the neuronal oscillator. Furthermore, it is difficult to
 derive the evolution of the transient ISI density from the current 
setting of the stochastic Poincar\'{e} oscillator, and thus we did 
not calculate the interspike interval serial correlation coefficient 
analytically.  Instead, 
we  introduced the instantaneous stochastic rotation number in 
this study and analyzed the corresponding components. In this way, 
one can quantify the effect of the past activity of neuronal oscillators. 
It is also a topic for future study to derive the transient counterparts 
of the conventional statistics.

Equations (\ref{eq:FullStoRotDecomp}) and
(\ref{eq:ReducedStoRotDecomp}) offer a way of determining the
components of the current instantaneous stochastic rotation number
experimentally. As an example, consider the instantaneous stochastic
rotation numbers corresponding to five impulses with fixed impulse
amplitudes, i.e., the impulses are characterized by four
interimpulse intervals $I_1$, $I_2$, $I_3$, and $I_4$ (Figure
\ref{Figure_label9} and Eq.~(\ref{eq:StoI4I3I2I1})). Consider first
the instantaneous stochastic rotation number corresponding to the
last interimpulse interval. The equation that determines the
stochastic rotation number at the last interimpulse interval is
Eq.~(\ref{eq:StoI4I3I2I1}), and to begin, we add five impulses with
interimpulse intervals of $I_1$, $I_2$, $I_3$, and $I_4$ and an 
inter-trial interval repeatedly
to a spontaneously firing neuron with some noise to measure the
firing rate during the interimpulse interval $I_4$,
$\Omega_{K,\epsilon,A,I_4}$ (Figure \ref{Figure_label9}A). After a
recovery period, we add a sufficient number of impulses with an
interimpulse interval of $I_1$ to the neuron to achieve the
corresponding steady state $h_{K,\epsilon,A,I_1}^*$ and then add
four impulses with interimpulse intervals $I_2$, $I_3$, and $I_4$ 
and an inter-trial interval to the same neuron (Figure \ref{Figure_label9}B). 
After repeating this procedure, the measured firing rate during the interimpulse interval
$I_4$ ($\Omega_{K,\epsilon,A,I_4,B}$) in Figure \ref{Figure_label9}B
is equal to the summation of the first, second, and third terms in
Eq.~(\ref{eq:StoI4I3I2I1}), since
$h_1(r,\phi)=h_{K,\epsilon,A,I_1}^*$, and this leads to
$Q_{K,\epsilon,A,I_1}h_1(r,\phi)=0$ (see the definition of
$Q_{K,\epsilon,A,I}$ in Eq.~(\ref{eq:SpecDecomp})). That is, the
fourth component in Eq.~(\ref{eq:StoI4I3I2I1}) is equal to zero.
Thus, $\Omega_{K,\epsilon,A,I_4}-\Omega_{K,\epsilon,A,I_4,B}$ is
equal to the fourth term in Eq.~(\ref{eq:StoI4I3I2I1}). After a
recovery period, we then add a sufficient number of impulses with an
interval $I_2$ to the same neuron to achieve the corresponding
steady state $h_{K,\epsilon,A,I_2}^*$ and then add three impulses
with interimpulse intervals of $I_3$ and $I_4$ and an inter-trial interval. 
After repeating this stimulation, the firing rate during the interimpulse interval $I_4$
($\Omega_{K,\epsilon,A,I_4,C}$) in Figure \ref{Figure_label9}C is
equal to the summation the first and second terms in
Eq.~(\ref{eq:StoI4I3I2I1}). Thus,
$\Omega_{K,\epsilon,A,I_4,B}-\Omega_{K,\epsilon,A,I_4,C}$ is equal
to the third term in Eq.~(\ref{eq:StoI4I3I2I1}). Again, after a
recovery period, impulses with an interval $I_3$ are added to the
same neuron to achieve the corresponding steady state
$h_{K,\epsilon,A,I_3}^*$, and two impulses with an interimpulse
interval of $I_4$ are added to the same neuron. After repeating this
stimulation with an inter-trial interval, the firing rate during $I_4$
($\Omega_{K,\epsilon,A,I_4,D}$) in Figure \ref{Figure_label9}D is
equal to the first term in Eq.~(\ref{eq:StoI4I3I2I1}). Thus,
$\Omega_{K,\epsilon,A,I_4,C}-\Omega_{K,\epsilon,A,I_4,D}$ is equal
to the second term in Eq.~(\ref{eq:StoI4I3I2I1}). In this way, one
can experimentally decompose the current instantaneous stochastic
rotation number.
The intrinsic noise strength of a neuron depends on the number 
of ion channels. The dynamic-clamp technique is one possible way 
to change the channel noise experimentally \cite{Dorval2005}. 
Combining this technique and the above-mentioned stimulus makes 
it possible to decompose the current instantaneous stochastic 
rotation number and examine the effect of the intrinsic noise 
strength on the decomposition.  

An increase in the noise
strength smoothed and decreased the variation in the components of
the instantaneous stochastic rotation number, which means that the
dependence on the instantaneous stochastic rotation number may
decrease as the noise strength increases. However, temporal coding
is not possible in this case because the larger noise makes the
firing time inaccurate. Furthermore, the experimental results of Perkel et al.
\cite{Perkel1964} revealed phase locking
in pacemaker neurons, and to consider the larger impulse amplitude,
it is necessary to understand the global picture of the response of
spontaneously firing neurons.

In terms of the dynamics of a spontaneously firing neuron to
response to time-varying impulses, it is necessary to investigate
the statistical behavior of the neuron in response to impulses with
a constant amplitude and interimpulse interval. Thus, it is
necessary to develop a method of constructing the SPTO or its analog
experimentally. It appears to be clear that if a spontaneously
firing neuron is modeled by stochastic differential equations, this
problem reduces to constructing the SPTO using the given equations.
The construction of the SPTO from stochastic differential equations
is not model dependent, since the small disturbance asymptotic
theory is general. Thus, one can construct the SPTO for other
impulse-driven biological or nonlinear oscillators.

Finally, for the case in which the instantaneous stochastic rotation
number depends on the past activity, it is difficult to encode information in
 the spike pattern since the spike generation also depends on the past activity. 
Our method can evaluates the dependence of 
the current state of a neuronal oscillator on the past 
activity. The amount of the past activity depended on the input parameters,
 the relaxation rate, noise strength. To evaluate the past dependency of various 
neurons and their models using our method or its extension would offer a way to 
get an insight into the information carrier in nervous systems.

\begin{acknowledgments}
The author thanks Professor A. Takahashi for providing
preprints. This work was supported by the Japan Science and Technology Agency,
Precursory Research for Embryonic Science and Technology Program.
\end{acknowledgments}

\bibliography{PREGlobalDynamics20130726}

\providecommand{\noopsort}[1]{}\providecommand{\singleletter}[1]{#1}%
\begin{thebibliography}{10}%
\makeatletter
\providecommand \@ifxundefined [1]{%
 \ifx #1\undefined \expandafter \@firstoftwo
 \else \expandafter \@secondoftwo
\fi
}%
\providecommand \@ifnum [1]{%
 \ifnum #1\expandafter \@firstoftwo
 \else \expandafter \@secondoftwo
\fi
}%
\providecommand \enquote [1]{``#1''}%
\providecommand \bibnamefont  [1]{#1}%
\providecommand \bibfnamefont [1]{#1}%
\providecommand \citenamefont [1]{#1}%
\providecommand\href[0]{\@sanitize\@href}%
\providecommand\@href[1]{\endgroup\@@startlink{#1}\endgroup\@@href}%
\providecommand\@@href[1]{#1\@@endlink}%
\providecommand \@sanitize [0]{\begingroup\catcode`\&12\catcode`\#12\relax}%
\@ifxundefined \pdfoutput {\@firstoftwo}{%
 \@ifnum{\z@=\pdfoutput}{\@firstoftwo}{\@secondoftwo}%
}{%
 \providecommand\@@startlink[1]{\leavevmode\special{html:<a href="#1">}}%
 \providecommand\@@endlink[0]{\special{html:</a>}}%
}{%
 \providecommand\@@startlink[1]{%
  \leavevmode
  \pdfstartlink
   attr{/Border[0 0 1 ]/H/I/C[0 1 1]}%
   user{/Subtype/Link/A<</Type/Action/S/URI/URI(#1)>>}%
  \relax
 }%
 \providecommand\@@endlink[0]{\pdfendlink}%
}%
\providecommand \url  [0]{\begingroup\@sanitize \@url }%
\providecommand \@url [1]{\endgroup\@href {#1}{\urlprefix}}%
\providecommand \urlprefix [0]{URL }%
\providecommand \Eprint[0]{\href }%
\@ifxundefined \urlstyle {%
  \providecommand \doi [1]{doi:\discretionary{}{}{}#1}%
}{%
  \providecommand \doi [0]{doi:\discretionary{}{}{}\begingroup
  \urlstyle{rm}\Url }%
}%
\providecommand \doibase [0]{http://dx.doi.org/}%
\providecommand \Doi[1]{\href{\doibase#1}}%
\providecommand \bibAnnote [3]{%
  \BibitemShut{#1}%
  \begin{quotation}\noindent
    \textsc{Key:}\ #2\\\textsc{Annotation:}\ #3%
  \end{quotation}%
}%
\providecommand \bibAnnoteFile [2]{%
  \IfFileExists{#2}{\bibAnnote {#1} {#2} {\input{#2}}}{}%
}%
\providecommand \typeout [0]{\immediate \write \m@ne }%
\providecommand \selectlanguage [0]{\@gobble}%
\providecommand \bibinfo [0]{\@secondoftwo}%
\providecommand \bibfield [0]{\@secondoftwo}%
\providecommand \translation [1]{[#1]}%
\providecommand \BibitemOpen[0]{}%
\providecommand \bibitemStop [0]{}%
\providecommand \bibitemNoStop [0]{.\EOS\space}%
\providecommand \EOS [0]{\spacefactor3000\relax}%
\providecommand \BibitemShut [1]{\csname bibitem#1\endcsname}%
\bibitem{RiekeWarlandSteveninckBialek1999}%
  \BibitemOpen
  \bibfield{author}{%
  \bibinfo {author} {\bibfnamefont{F.}~\bibnamefont{Rieke}}, \bibinfo {author}
  {\bibfnamefont{D.}~\bibnamefont{Warland}}, \bibinfo {author}
  {\bibfnamefont{R.}~\bibnamefont{de~de Ruyter~van Steveninck}},\ and\ \bibinfo
  {author} {\bibfnamefont{W.}~\bibnamefont{Bialek}},\ }%
  \emph{\bibinfo {title} {Spikes: Exploring the Neural Code (Computational
  Neuroscience)}}\ (\bibinfo {publisher} {A Bradford Book},\ \bibinfo {year}
  {1999})%
  \bibAnnoteFile{NoStop}{RiekeWarlandSteveninckBialek1999}%
\bibitem{SegundoStiberAltshulerVibert1994}%
  \BibitemOpen
  \bibfield{author}{%
  \bibinfo {author} {\bibfnamefont{J.~P.}\ \bibnamefont{Segundo}}, \bibinfo
  {author} {\bibfnamefont{M.}~\bibnamefont{Stiber}}, \bibinfo {author}
  {\bibfnamefont{E.}~\bibnamefont{Altshuler}},\ and\ \bibinfo {author}
  {\bibfnamefont{J.~F.}\ \bibnamefont{Vibert}},\ }%
  \bibfield{journal}{%
  \bibinfo {journal} {Neuroscience}\ }%
  \textbf{\bibinfo {volume} {62}},\ \bibinfo {pages} {459} (\bibinfo {year}
  {1994})%
  \bibAnnoteFile{NoStop}{SegundoStiberAltshulerVibert1994}%
\bibitem{StiberIeongSegundo1997}%
  \BibitemOpen
  \bibfield{author}{%
  \bibinfo {author} {\bibfnamefont{M.}~\bibnamefont{Stiber}}, \bibinfo {author}
  {\bibfnamefont{R.}~\bibnamefont{Ieong}},\ and\ \bibinfo {author}
  {\bibfnamefont{J.}~\bibnamefont{Segundo}},\ }%
  \bibfield{journal}{%
  \bibinfo {journal} {IEEE T. Neural Networ.}\ }%
  \textbf{\bibinfo {volume} {8}},\ \bibinfo {pages} {1379} (\bibinfo {year}
  {1997})%
  \bibAnnoteFile{NoStop}{StiberIeongSegundo1997}%
\bibitem{YamanobePakdamanNomuraSato1998}%
  \BibitemOpen
  \bibfield{author}{%
  \bibinfo {author} {\bibfnamefont{T.}~\bibnamefont{Yamanobe}}, \bibinfo
  {author} {\bibfnamefont{K.}~\bibnamefont{Pakdaman}}, \bibinfo {author}
  {\bibfnamefont{T.}~\bibnamefont{Nomura}},\ and\ \bibinfo {author}
  {\bibfnamefont{S.}~\bibnamefont{Sato}},\ }%
  \bibfield{journal}{%
  \bibinfo {journal} {Biosystems}\ }%
  \textbf{\bibinfo {volume} {48}},\ \bibinfo {pages} {287} (\bibinfo {year}
  {1998})%
  \bibAnnoteFile{NoStop}{YamanobePakdamanNomuraSato1998}%
\bibitem{Yamanobe2011}%
  \BibitemOpen
  \bibfield{author}{%
  \bibinfo {author} {\bibfnamefont{T.}~\bibnamefont{Yamanobe}},\ }%
  \bibfield{journal}{%
  \bibinfo {journal} {Phys. Rev. E Stat. Nonlin. Soft Matter Phys.}\ }%
  \textbf{\bibinfo {volume} {84}},\ \bibinfo {pages} {011924} (\bibinfo {year}
  {2011})%
  \bibAnnoteFile{NoStop}{Yamanobe2011}%
\bibitem{RabinovichBaronaSelverstonAbarbanel2006}%
  \BibitemOpen
  \bibfield{author}{%
  \bibinfo {author} {\bibfnamefont{M.~I.}\ \bibnamefont{Rabinovich}}, \bibinfo
  {author} {\bibfnamefont{P.}~\bibnamefont{Varona}}, \bibinfo {author}
  {\bibfnamefont{A.~I.}\ \bibnamefont{Selverston}},\ and\ \bibinfo {author}
  {\bibfnamefont{H.~D.}\ \bibnamefont{Abarbanel}},\ }%
  \bibfield{journal}{%
  \bibinfo {journal} {Rev. Mod. Phys.}\ }%
  \textbf{\bibinfo {volume} {78}},\ \bibinfo {pages} {1213} (\bibinfo {year}
  {2006})%
  \bibAnnoteFile{NoStop}{RabinovichBaronaSelverstonAbarbanel2006}%
\bibitem{RabinovichHuertaLaurent2010}%
  \BibitemOpen
  \bibfield{author}{%
  \bibinfo {author} {\bibfnamefont{M.}~\bibnamefont{Rabinovich}}, \bibinfo
  {author} {\bibfnamefont{R.}~\bibnamefont{Huerta}},\ and\ \bibinfo {author}
  {\bibfnamefont{G.}~\bibnamefont{Laurent}},\ }%
  \bibfield{journal}{%
  \bibinfo {journal} {Science}\ }%
  \textbf{\bibinfo {volume} {321}},\ \bibinfo {pages} {48} (\bibinfo {year}
  {2008})%
  \bibAnnoteFile{NoStop}{RabinovichHuertaLaurent2010}%
\bibitem{HoppensteadtIzhikevich1997}%
  \BibitemOpen
  \bibfield{author}{%
  \bibinfo {author} {\bibfnamefont{F.~C.}\ \bibnamefont{Hoppensteadt}}\ and\
  \bibinfo {author} {\bibfnamefont{E.~M.}\ \bibnamefont{Izhikevich}},\ }%
  \emph{\bibinfo {title} {Weakly Connected Neural Networks (Applied
  Mathematical Sciences)}}\ (\bibinfo {publisher} {Springer},\ \bibinfo {year}
  {1997})%
  \bibAnnoteFile{NoStop}{HoppensteadtIzhikevich1997}%
\bibitem{Izhikevich2010}%
  \BibitemOpen
  \bibfield{author}{%
  \bibinfo {author} {\bibfnamefont{E.~M.}\ \bibnamefont{Izhikevich}},\ }%
  \emph{\bibinfo {title} {Dynamical Systems in Neuroscience: The Geometry of
  Excitability and Bursting}}\ (\bibinfo {publisher} {The MIT Press},\ \bibinfo
  {year} {2010})%
  \bibAnnoteFile{NoStop}{Izhikevich2010}%
\bibitem{GuevaraGlassShrier1981}%
  \BibitemOpen
  \bibfield{author}{%
  \bibinfo {author} {\bibfnamefont{M.~R.}\ \bibnamefont{Guevara}}, \bibinfo
  {author} {\bibfnamefont{L.}~\bibnamefont{Glass}},\ and\ \bibinfo {author}
  {\bibfnamefont{A.}~\bibnamefont{Shrier}},\ }%
  \bibfield{journal}{%
  \bibinfo {journal} {Science}\ }%
  \textbf{\bibinfo {volume} {214}},\ \bibinfo {pages} {1350} (\bibinfo {year}
  {1981})%
  \bibAnnoteFile{NoStop}{GuevaraGlassShrier1981}%
\bibitem{GuevaraShrierGlassPerez1983}%
  \BibitemOpen
  \bibfield{author}{%
  \bibinfo {author} {\bibfnamefont{L.}~\bibnamefont{Glass}}, \bibinfo {author}
  {\bibfnamefont{M.~R.}\ \bibnamefont{Guevara}}, \bibinfo {author}
  {\bibfnamefont{A.}~\bibnamefont{Shrier}},\ and\ \bibinfo {author}
  {\bibfnamefont{R.}~\bibnamefont{Perez}},\ }%
  \bibfield{journal}{%
  \Doi{10.1016/0167-2789(83)90119-7}{\bibinfo {journal} {Physica D: Nonlinear
  Phenomena}}\ }%
  \textbf{\bibinfo {volume} {7}},\ \bibinfo {pages} {89 } (\bibinfo {year}
  {1983})%
  \bibAnnoteFile{NoStop}{GuevaraShrierGlassPerez1983}%
\bibitem{SmithEllenbergerBallanyiRichterFeldman1991}%
  \BibitemOpen
  \bibfield{author}{%
  \bibinfo {author} {\bibfnamefont{J.~C.}\ \bibnamefont{Smith}}, \bibinfo
  {author} {\bibfnamefont{H.~H.}\ \bibnamefont{Ellenberger}}, \bibinfo {author}
  {\bibfnamefont{K.}~\bibnamefont{Ballanyi}}, \bibinfo {author}
  {\bibfnamefont{D.~W.}\ \bibnamefont{Richter}},\ and\ \bibinfo {author}
  {\bibfnamefont{J.~L.}\ \bibnamefont{Feldman}},\ }%
  \bibfield{journal}{%
  \bibinfo {journal} {Science}\ }%
  \textbf{\bibinfo {volume} {254}},\ \bibinfo {pages} {726} (\bibinfo {year}
  {1991})%
  \bibAnnoteFile{NoStop}{SmithEllenbergerBallanyiRichterFeldman1991}%
\bibitem{Johnson1994}%
  \BibitemOpen
  \bibfield{author}{%
  \bibinfo {author} {\bibfnamefont{S.~M.}\ \bibnamefont{Johnson}}, \bibinfo
  {author} {\bibfnamefont{J.~C.}\ \bibnamefont{Smith}}, \bibinfo {author}
  {\bibfnamefont{G.~D.}\ \bibnamefont{Funk}},\ and\ \bibinfo {author}
  {\bibfnamefont{J.~L.}\ \bibnamefont{Feldman}},\ }%
  \bibfield{journal}{%
  \bibinfo {journal} {J. Neurophysiol.}\ }%
  \textbf{\bibinfo {volume} {72}},\ \bibinfo {pages} {2598} (\bibinfo {year}
  {1994})%
  \bibAnnoteFile{NoStop}{Johnson1994}%
\bibitem{WiesenfeldColetStrogatz1998}%
  \BibitemOpen
  \bibfield{author}{%
  \bibinfo {author} {\bibfnamefont{K.}~\bibnamefont{Wiesenfeld}}, \bibinfo
  {author} {\bibfnamefont{P.}~\bibnamefont{Colet}},\ and\ \bibinfo {author}
  {\bibfnamefont{S.~H.}\ \bibnamefont{Strogatz}},\ }%
  \bibfield{journal}{%
  \bibinfo {journal} {Phys. Rev. E Stat. Nonlin. Soft Matter Phys.}\ }%
  \textbf{\bibinfo {volume} {57}},\ \bibinfo {pages} {1563} (\bibinfo {year}
  {1998})%
  \bibAnnoteFile{NoStop}{WiesenfeldColetStrogatz1998}%
\bibitem{Jin1997}%
  \BibitemOpen
  \bibfield{author}{%
  \bibinfo {author} {\bibfnamefont{F.-F.}\ \bibnamefont{Jin}},\ }%
  \bibfield{journal}{%
  \bibinfo {journal} {J. Atmos. Sci.}\ }%
  \textbf{\bibinfo {volume} {54}},\ \bibinfo {pages} {811} (\bibinfo {year}
  {1997})%
  \bibAnnoteFile{NoStop}{Jin1997}%
\bibitem{Kuramoto2003}%
  \BibitemOpen
  \bibfield{author}{%
  \bibinfo {author} {\bibfnamefont{Y.}~\bibnamefont{Kuramoto}},\ }%
  \emph{\bibinfo {title} {Chemical Oscillations, Waves, and Turbulence (Dover
  Books on Chemistry)}}\ (\bibinfo {publisher} {Dover Publications},\ \bibinfo
  {year} {2003})%
  \bibAnnoteFile{NoStop}{Kuramoto2003}%
\bibitem{Winfree2010}%
  \BibitemOpen
  \bibfield{author}{%
  \bibinfo {author} {\bibfnamefont{A.~T.}\ \bibnamefont{Winfree}},\ }%
  \emph{\bibinfo {title} {The Geometry of Biological Time (Interdisciplinary
  Applied Mathematics)}}\ (\bibinfo {publisher} {Springer},\ \bibinfo {year}
  {2010})%
  \bibAnnoteFile{NoStop}{Winfree2010}%
\bibitem{Hoppensteadt1982}%
  \BibitemOpen
  \bibfield{author}{%
  \bibinfo {author} {\bibfnamefont{F.~C.}\ \bibnamefont{Hoppensteadt}}\ and\
  \bibinfo {author} {\bibfnamefont{J.~P.}\ \bibnamefont{Keener}},\ }%
  \bibfield{journal}{%
  \bibinfo {journal} {J. Math. Biol.}\ }%
  \textbf{\bibinfo {volume} {15}},\ \bibinfo {pages} {339} (\bibinfo {year}
  {1982})%
  \bibAnnoteFile{NoStop}{Hoppensteadt1982}%
\bibitem{GlassMackey1988}%
  \BibitemOpen
  \bibfield{author}{%
  \bibinfo {author} {\bibfnamefont{L.}~\bibnamefont{Glass}}\ and\ \bibinfo
  {author} {\bibfnamefont{M.~C.}\ \bibnamefont{Mackey}},\ }%
  \emph{\bibinfo {title} {From Clocks to Chaos}}\ (\bibinfo {publisher}
  {Princeton University Press},\ \bibinfo {year} {1988})%
  \bibAnnoteFile{NoStop}{GlassMackey1988}%
\bibitem{GlassSun1994}%
  \BibitemOpen
  \bibfield{author}{%
  \bibinfo {author} {\bibfnamefont{L.}~\bibnamefont{Glass}}\ and\ \bibinfo
  {author} {\bibfnamefont{J.}~\bibnamefont{Sun}},\ }%
  \bibfield{journal}{%
  \bibinfo {journal} {Phys. Rev. E Stat. Nonlin. Soft Matter Phys.}\ }%
  \textbf{\bibinfo {volume} {50}},\ \bibinfo {pages} {5077} (\bibinfo {year}
  {1994})%
  \bibAnnoteFile{NoStop}{GlassSun1994}%
\bibitem{StiberPakdamanVibertBoussardSegundoNomuraSatoDoi1997}%
  \BibitemOpen
  \bibfield{author}{%
  \bibinfo {author} {\bibfnamefont{M.}~\bibnamefont{Stiber}}, \bibinfo {author}
  {\bibfnamefont{K.}~\bibnamefont{Pakdaman}}, \bibinfo {author}
  {\bibfnamefont{J.~F.}\ \bibnamefont{Vibert}}, \bibinfo {author}
  {\bibfnamefont{E.}~\bibnamefont{Boussard}}, \bibinfo {author}
  {\bibfnamefont{J.~P.}\ \bibnamefont{Segundo}}, \bibinfo {author}
  {\bibfnamefont{T.}~\bibnamefont{Nomura}}, \bibinfo {author}
  {\bibfnamefont{S.}~\bibnamefont{Sato}},\ and\ \bibinfo {author}
  {\bibfnamefont{S.}~\bibnamefont{Doi}},\ }%
  \bibfield{journal}{%
  \bibinfo {journal} {Biosystems}\ }%
  \textbf{\bibinfo {volume} {40}},\ \bibinfo {pages} {177} (\bibinfo {year}
  {1997})%
  \bibAnnoteFile{NoStop}{StiberPakdamanVibertBoussardSegundoNomuraSatoDoi1997}%
\bibitem{NomuraSatoDoiSegundoStiber1994}%
  \BibitemOpen
  \bibfield{author}{%
  \bibinfo {author} {\bibfnamefont{T.}~\bibnamefont{Nomura}}, \bibinfo {author}
  {\bibfnamefont{S.}~\bibnamefont{Sato}}, \bibinfo {author}
  {\bibfnamefont{S.}~\bibnamefont{Doi}}, \bibinfo {author}
  {\bibfnamefont{J.~P.}\ \bibnamefont{Segundo}},\ and\ \bibinfo {author}
  {\bibfnamefont{M.~D.}\ \bibnamefont{Stiber}},\ }%
  \bibfield{journal}{%
  \bibinfo {journal} {Biol. Cybern.}\ }%
  \textbf{\bibinfo {volume} {72}},\ \bibinfo {pages} {93} (\bibinfo {year}
  {1994})%
  \bibAnnoteFile{NoStop}{NomuraSatoDoiSegundoStiber1994}%
\bibitem{YamanobePakdaman2002}%
  \BibitemOpen
  \bibfield{author}{%
  \bibinfo {author} {\bibfnamefont{T.}~\bibnamefont{Yamanobe}}\ and\ \bibinfo
  {author} {\bibfnamefont{K.}~\bibnamefont{Pakdaman}},\ }%
  \bibfield{journal}{%
  \bibinfo {journal} {Biol. Cybern.}\ }%
  \textbf{\bibinfo {volume} {86}},\ \bibinfo {pages} {155} (\bibinfo {year}
  {2002})%
  \bibAnnoteFile{NoStop}{YamanobePakdaman2002}%
\bibitem{PakdamanMestivier2001}%
  \BibitemOpen
  \bibfield{author}{%
  \bibinfo {author} {\bibfnamefont{K.}~\bibnamefont{Pakdaman}}\ and\ \bibinfo
  {author} {\bibfnamefont{D.}~\bibnamefont{Mestivier}},\ }%
  \bibfield{journal}{%
  \bibinfo {journal} {Phys. Rev. E Stat. Nonlin. Soft Matter Phys.}\ }%
  \textbf{\bibinfo {volume} {64}},\ \bibinfo {pages} {030901} (\bibinfo {year}
  {2001})%
  \bibAnnoteFile{NoStop}{PakdamanMestivier2001}%
\bibitem{CroisierGuevaraDauby2009}%
  \BibitemOpen
  \bibfield{author}{%
  \bibinfo {author} {\bibfnamefont{H.}~\bibnamefont{Croisier}}, \bibinfo
  {author} {\bibfnamefont{M.~R.}\ \bibnamefont{Guevara}},\ and\ \bibinfo
  {author} {\bibfnamefont{P.~C.}\ \bibnamefont{Dauby}},\ }%
  \bibfield{journal}{%
  \bibinfo {journal} {Phys. Rev. E: Stat., Nonlinear, Soft Matter Phys.}\ }%
  \textbf{\bibinfo {volume} {79}},\ \bibinfo {pages} {016209} (\bibinfo {year}
  {2009})%
  \bibAnnoteFile{NoStop}{CroisierGuevaraDauby2009}%
\bibitem{WhiteRubinsteinKay2000}%
  \BibitemOpen
  \bibfield{author}{%
  \bibinfo {author} {\bibfnamefont{J.~A.}\ \bibnamefont{White}}, \bibinfo
  {author} {\bibfnamefont{J.~T.}\ \bibnamefont{Rubinstein}},\ and\ \bibinfo
  {author} {\bibfnamefont{A.~R.}\ \bibnamefont{Kay}},\ }%
  \bibfield{journal}{%
  \bibinfo {journal} {Trends Neurosci.}\ }%
  \textbf{\bibinfo {volume} {23}},\ \bibinfo {pages} {131} (\bibinfo {year}
  {2000})%
  \bibAnnoteFile{NoStop}{WhiteRubinsteinKay2000}%
\bibitem{KunitomoTakahashi2003}%
  \BibitemOpen
  \bibfield{author}{%
  \bibinfo {author} {\bibfnamefont{N.}~\bibnamefont{Kunitomo}}\ and\ \bibinfo
  {author} {\bibfnamefont{A.}~\bibnamefont{Takahashi}},\ }%
  \bibfield{journal}{%
  \bibinfo {journal} {Ann. Appl. Probab.}\ }%
  \textbf{\bibinfo {volume} {13}},\ \bibinfo {pages} {914} (\bibinfo {year}
  {2003})%
  \bibAnnoteFile{NoStop}{KunitomoTakahashi2003}%
\bibitem{TakahashiTakeharaToda2011}%
  \BibitemOpen
  \bibfield{author}{%
  \bibinfo {author} {\bibfnamefont{A.}~\bibnamefont{Takahashi}}, \bibinfo
  {author} {\bibfnamefont{K.}~\bibnamefont{Takehara}},\ and\ \bibinfo {author}
  {\bibfnamefont{M.}~\bibnamefont{Toda}},\ }%
  \bibfield{journal}{%
  \bibinfo {journal} {CARF-F-242}}%
   (\bibinfo {year} {2011})%
  \bibAnnoteFile{NoStop}{TakahashiTakeharaToda2011}%
\bibitem{Guevara1982}%
  \BibitemOpen
  \bibfield{author}{%
  \bibinfo {author} {\bibfnamefont{M.~R.}\ \bibnamefont{Guevara}}\ and\
  \bibinfo {author} {\bibfnamefont{L.}~\bibnamefont{Glass}},\ }%
  \bibfield{journal}{%
  \bibinfo {journal} {J. Math. Biol.}\ }%
  \textbf{\bibinfo {volume} {14}},\ \bibinfo {pages} {1} (\bibinfo {year}
  {1982})%
  \bibAnnoteFile{NoStop}{Guevara1982}%
\bibitem{Keener1984}%
  \BibitemOpen
  \bibfield{author}{%
  \bibinfo {author} {\bibfnamefont{J.~P.}\ \bibnamefont{Keener}}\ and\ \bibinfo
  {author} {\bibfnamefont{L.}~\bibnamefont{Glass}},\ }%
  \bibfield{journal}{%
  \bibinfo {journal} {J. Math. Biol.}\ }%
  \textbf{\bibinfo {volume} {21}},\ \bibinfo {pages} {175} (\bibinfo {year}
  {1984})%
  \bibAnnoteFile{NoStop}{Keener1984}%
\bibitem{Glass2001}%
  \BibitemOpen
  \bibfield{author}{%
  \bibinfo {author} {\bibfnamefont{L.}~\bibnamefont{Glass}},\ }%
  \bibfield{journal}{%
  \bibinfo {journal} {Nature}\ }%
  \textbf{\bibinfo {volume} {410}},\ \bibinfo {pages} {277} (\bibinfo {year}
  {2001})%
  \bibAnnoteFile{NoStop}{Glass2001}%
\bibitem{Tateno2007a}%
  \BibitemOpen
  \bibfield{author}{%
  \bibinfo {author} {\bibfnamefont{T.}~\bibnamefont{Tateno}}\ and\ \bibinfo
  {author} {\bibfnamefont{H.~P.~C.}\ \bibnamefont{Robinson}},\ }%
  \bibfield{journal}{%
  \bibinfo {journal} {Biophys. J.}\ }%
  \textbf{\bibinfo {volume} {92}},\ \bibinfo {pages} {683} (\bibinfo {year}
  {2007})%
  \bibAnnoteFile{NoStop}{Tateno2007a}%
\bibitem{YoshimuraArai2008}%
  \BibitemOpen
  \bibfield{author}{%
  \bibinfo {author} {\bibfnamefont{K.}~\bibnamefont{Yoshimura}}\ and\ \bibinfo
  {author} {\bibfnamefont{K.}~\bibnamefont{Arai}},\ }%
  \bibfield{journal}{%
  \bibinfo {journal} {Phys. Rev. Lett.}\ }%
  \textbf{\bibinfo {volume} {101}},\ \bibinfo {pages} {154101} (\bibinfo {year}
  {2008})%
  \bibAnnoteFile{NoStop}{YoshimuraArai2008}%
\bibitem{Gantmacher1959}%
  \BibitemOpen
  \bibfield{author}{%
  \bibinfo {author} {\bibfnamefont{F.~R.}\ \bibnamefont{Gantmacher}},\ }%
  \emph{\bibinfo {title} {The Theory of Matrices, Vol. II}}\ (\bibinfo
  {publisher} {Chelsea Publishing Company},\ \bibinfo {year} {1959})%
  \bibAnnoteFile{NoStop}{Gantmacher1959}%
\bibitem{LasotaMackey1998}%
  \BibitemOpen
  \bibfield{author}{%
  \bibinfo {author} {\bibfnamefont{A.}~\bibnamefont{Lasota}}\ and\ \bibinfo
  {author} {\bibfnamefont{M.~C.}\ \bibnamefont{Mackey}},\ }%
  \emph{\bibinfo {title} {Chaos, Fractals, and Noise: Stochastic Aspects of
  Dynamics (Applied Mathematical Sciences)}}\ (\bibinfo {publisher}
  {Springer},\ \bibinfo {year} {1998})%
  \bibAnnoteFile{NoStop}{LasotaMackey1998}%
\bibitem{DingZhou2009}%
  \BibitemOpen
  \bibfield{author}{%
  \bibinfo {author} {\bibfnamefont{J.}~\bibnamefont{Ding}}\ and\ \bibinfo
  {author} {\bibfnamefont{A.}~\bibnamefont{Zhou}},\ }%
  \emph{\bibinfo {title} {Nonnegative Matrices, Positive Operators, and
  Applications}}\ (\bibinfo {publisher} {World Scientific Pub Co Inc},\
  \bibinfo {year} {2009})%
  \bibAnnoteFile{NoStop}{DingZhou2009}%
\bibitem{DoiInoueKumagai1998}%
  \BibitemOpen
  \bibfield{author}{%
  \bibinfo {author} {\bibfnamefont{S.}~\bibnamefont{Doi}}, \bibinfo {author}
  {\bibfnamefont{J.}~\bibnamefont{Inoue}},\ and\ \bibinfo {author}
  {\bibfnamefont{S.}~\bibnamefont{Kumagai}},\ }%
  \bibfield{journal}{%
  \bibinfo {journal} {J. Stat. Phys.}\ }%
  \textbf{\bibinfo {volume} {90}},\ \bibinfo {pages} {1107} (\bibinfo {year}
  {1998})%
  \bibAnnoteFile{NoStop}{DoiInoueKumagai1998}%
\bibitem{TatenoDoiSatoRicciardi1995}%
  \BibitemOpen
  \bibfield{author}{%
  \bibinfo {author} {\bibfnamefont{T.}~\bibnamefont{Tateno}}, \bibinfo {author}
  {\bibfnamefont{S.}~\bibnamefont{Doi}}, \bibinfo {author}
  {\bibfnamefont{S.}~\bibnamefont{Sato}},\ and\ \bibinfo {author}
  {\bibfnamefont{L.~M.}\ \bibnamefont{Ricciardi}},\ }%
  \bibfield{journal}{%
  \bibinfo {journal} {J. Stat. Phys.}\ }%
  \textbf{\bibinfo {volume} {78}},\ \bibinfo {pages} {917} (\bibinfo {year}
  {1995})%
  \bibAnnoteFile{NoStop}{TatenoDoiSatoRicciardi1995}%
\bibitem{Tateno1998}%
  \BibitemOpen
  \bibfield{author}{%
  \bibinfo {author} {\bibfnamefont{T.}~\bibnamefont{Tateno}},\ }%
  \bibfield{journal}{%
  \bibinfo {journal} {J. Stat. Phys.}\ }%
  \textbf{\bibinfo {volume} {92}},\ \bibinfo {pages} {675} (\bibinfo {year}
  {1998})%
  \bibAnnoteFile{NoStop}{Tateno1998}%
\bibitem{InoueDoiKumagai2001}%
  \BibitemOpen
  \bibfield{author}{%
  \bibinfo {author} {\bibfnamefont{J.}~\bibnamefont{Inoue}}, \bibinfo {author}
  {\bibfnamefont{S.}~\bibnamefont{Doi}},\ and\ \bibinfo {author}
  {\bibfnamefont{S.}~\bibnamefont{Kumagai}},\ }%
  \bibfield{journal}{%
  \bibinfo {journal} {Phys. Rev. E Stat. Nonlin. Soft Matter Phys.}\ }%
  \textbf{\bibinfo {volume} {64}},\ \bibinfo {pages} {056219} (\bibinfo {year}
  {2001})%
  \bibAnnoteFile{NoStop}{InoueDoiKumagai2001}%
\bibitem{Tateno2002}%
  \BibitemOpen
  \bibfield{author}{%
  \bibinfo {author} {\bibfnamefont{T.}~\bibnamefont{Tateno}},\ }%
  \bibfield{journal}{%
  \bibinfo {journal} {Phys. Rev. E Stat. Nonlin. Soft Matter Phys.}\ }%
  \textbf{\bibinfo {volume} {65}},\ \bibinfo {pages} {021901} (\bibinfo {year}
  {2002})%
  \bibAnnoteFile{NoStop}{Tateno2002}%
\bibitem{Arnold2008}%
  \BibitemOpen
  \bibfield{author}{%
  \bibinfo {author} {\bibfnamefont{L.}~\bibnamefont{Arnold}},\ }%
  \emph{\bibinfo {title} {Random Dynamical Systems (Springer Monographs in
  Mathematics)}}\ (\bibinfo {publisher} {Springer},\ \bibinfo {year} {2008})%
  \bibAnnoteFile{NoStop}{Arnold2008}%
\bibitem{BorisyukRassoulAgha}%
  \BibitemOpen
  \bibfield{author}{%
  \bibinfo {author} {\bibfnamefont{A.}~\bibnamefont{Borisyuk}}\ and\ \bibinfo
  {author} {\bibfnamefont{F.}~\bibnamefont{Rassoul-Agha}},\ }%
  \bibfield{journal}{%
  \bibinfo {journal} {SIAM Appl. Dyn. Syst.}}%
   (\bibinfo {year} {To appear})%
  \bibAnnoteFile{NoStop}{BorisyukRassoulAgha}%
\bibitem{Seneta2007}%
  \BibitemOpen
  \bibfield{author}{%
  \bibinfo {author} {\bibfnamefont{E.}~\bibnamefont{Seneta}},\ }%
  \emph{\bibinfo {title} {Non-negative Matrices and {Markov} Chains (Springer
  Series in Statistics)}}\ (\bibinfo {publisher} {Springer},\ \bibinfo {year}
  {2007})%
  \bibAnnoteFile{NoStop}{Seneta2007}%
\bibitem{IpsenSelee2011}%
  \BibitemOpen
  \bibfield{author}{%
  \bibinfo {author} {\bibfnamefont{I.~C.}\ \bibnamefont{Ipsen}}\ and\ \bibinfo
  {author} {\bibfnamefont{T.~M.}\ \bibnamefont{Selee}},\ }%
  \bibfield{journal}{%
  \bibinfo {journal} {SIAM J. Matrix Anal. A.}\ }%
  \textbf{\bibinfo {volume} {32}},\ \bibinfo {pages} {153} (\bibinfo {year}
  {2011})%
  \bibAnnoteFile{NoStop}{IpsenSelee2011}%
\bibitem{NesseClarkBressloff2007}%
  \BibitemOpen
  \bibfield{author}{%
  \bibinfo {author} {\bibfnamefont{W.~H.}\ \bibnamefont{Nesse}}, \bibinfo
  {author} {\bibfnamefont{G.~A.}\ \bibnamefont{Clark}},\ and\ \bibinfo {author}
  {\bibfnamefont{P.~C.}\ \bibnamefont{Bressloff}},\ }%
  \bibfield{journal}{%
  \bibinfo {journal} {Phys. Rev. E Stat. Nonlin. Soft Matter Phys.}\ }%
  \textbf{\bibinfo {volume} {75}},\ \bibinfo {pages} {031912} (\bibinfo {year}
  {2007})%
  \bibAnnoteFile{NoStop}{NesseClarkBressloff2007}%
\bibitem{VideoS1}%
  \BibitemOpen
  \bibfield{author}{%
  \bibinfo {author} {\bibnamefont{{Video S1}}},\ }%
  \bibinfo {note} {{Six movies showing the evolution of the density and
  corresponding transient components for $K=0.25$, $1$, and $\infty$ using the
  parameters given in Figure 3. The impulse amplitude was zero and the
  interimpulse interval \textit{I} (the time from the initial density as
  explained in Figure 3) was changed from 0.3 to 5.25 in steps of 0.05 in all
  movies. The time is shown at the top of each density or transient component
  movie. The name of each movie indicates the corresponding plot in Figure 3,
  and ``D'' indicates the density and ``T'' denotes the transient component.
  For example, VideoS1A-D.wmv is a movie of the density shown in Figure 3A
  (left panel) and VideoS1A-T.wmv is a movie of the transient component shown
  in Figure 3A (middle panel). The color bar shows the probability density or
  the corresponding transient component.}}%
  \bibAnnoteFile{Stop}{VideoS1}%
\bibitem{TatenoJimbo2000}%
  \BibitemOpen
  \bibfield{author}{%
  \bibinfo {author} {\bibfnamefont{T.}~\bibnamefont{Tateno}}\ and\ \bibinfo
  {author} {\bibfnamefont{Y.}~\bibnamefont{Jimbo}},\ }%
  \bibfield{journal}{%
  \bibinfo {journal} {Phys. Lett. A}\ }%
  \textbf{\bibinfo {volume} {271}},\ \bibinfo {pages} {227} (\bibinfo {year}
  {2000})%
  \bibAnnoteFile{NoStop}{TatenoJimbo2000}%
\bibitem{VideoS2}%
  \BibitemOpen
  \bibfield{author}{%
  \bibinfo {author} {\bibnamefont{{Video S2}}},\ }%
  \bibinfo {note} {{Dependence of the invariant density on the input rate for
  $K=1$, $\epsilon=0.3$, and $A=0.95$. The input rate was changed from 0.5 to
  3.33 in steps of 0.005. The input rate is shown at the top of each invariant
  density. Color bar indicates the probability density of the invariant
  density.}}%
  \bibAnnoteFile{Stop}{VideoS2}%
\bibitem{Parzen1962}%
  \BibitemOpen
  \bibfield{author}{%
  \bibinfo {author} {\bibfnamefont{E.}~\bibnamefont{Parzen}},\ }%
  \bibfield{journal}{%
  \bibinfo {journal} {Ann. Math. Stat.}\ }%
  \textbf{\bibinfo {volume} {33}},\ \bibinfo {pages} {1065} (\bibinfo {year}
  {1962})%
  \bibAnnoteFile{NoStop}{Parzen1962}%
\bibitem{Rosenblatt1956}%
  \BibitemOpen
  \bibfield{author}{%
  \bibinfo {author} {\bibfnamefont{M.}~\bibnamefont{Rosenblatt}},\ }%
  \bibfield{journal}{%
  \bibinfo {journal} {Ann. Math. Stat.}\ }%
  \textbf{\bibinfo {volume} {27}},\ \bibinfo {pages} {832} (\bibinfo {year}
  {1956})%
  \bibAnnoteFile{NoStop}{Rosenblatt1956}%
\bibitem{Sanderson1980}%
  \BibitemOpen
  \bibfield{author}{%
  \bibinfo {author} {\bibfnamefont{A.~C.}\ \bibnamefont{Sanderson}},\ }%
  \bibfield{journal}{%
  \bibinfo {journal} {IEEE T. Bio-med. Eng.}\ }%
  \textbf{\bibinfo {volume} {27}},\ \bibinfo {pages} {271} (\bibinfo {year}
  {1980})%
  \bibAnnoteFile{NoStop}{Sanderson1980}%
\bibitem{Richmond1990a}%
  \BibitemOpen
  \bibfield{author}{%
  \bibinfo {author} {\bibfnamefont{B.~J.}\ \bibnamefont{Richmond}}, \bibinfo
  {author} {\bibfnamefont{L.~M.}\ \bibnamefont{Optican}},\ and\ \bibinfo
  {author} {\bibfnamefont{H.}~\bibnamefont{Spitzer}},\ }%
  \bibfield{journal}{%
  \bibinfo {journal} {J. Neurophysiol.}\ }%
  \textbf{\bibinfo {volume} {64}},\ \bibinfo {pages} {351} (\bibinfo {year}
  {1990})%
  \bibAnnoteFile{NoStop}{Richmond1990a}%
\bibitem{Nawrot1999}%
  \BibitemOpen
  \bibfield{author}{%
  \bibinfo {author} {\bibfnamefont{M.}~\bibnamefont{Nawrot}}, \bibinfo {author}
  {\bibfnamefont{A.}~\bibnamefont{Aertsen}},\ and\ \bibinfo {author}
  {\bibfnamefont{S.}~\bibnamefont{Rotter}},\ }%
  \bibfield{journal}{%
  \bibinfo {journal} {J. Neurosci. Meth.}\ }%
  \textbf{\bibinfo {volume} {94}},\ \bibinfo {pages} {81} (\bibinfo {year}
  {1999})%
  \bibAnnoteFile{NoStop}{Nawrot1999}%
\bibitem{Shimazaki2010}%
  \BibitemOpen
  \bibfield{author}{%
  \bibinfo {author} {\bibfnamefont{H.}~\bibnamefont{Shimazaki}}\ and\ \bibinfo
  {author} {\bibfnamefont{S.}~\bibnamefont{Shinomoto}},\ }%
  \bibfield{journal}{%
  \bibinfo {journal} {J. Comput. Neurosci.}\ }%
  \textbf{\bibinfo {volume} {29}},\ \bibinfo {pages} {171} (\bibinfo {year}
  {2010})%
  \bibAnnoteFile{NoStop}{Shimazaki2010}%
\bibitem{SegundoVibertStiberHanneton1995a}%
  \BibitemOpen
  \bibfield{author}{%
  \bibinfo {author} {\bibfnamefont{J.~P.}\ \bibnamefont{Segundo}}, \bibinfo
  {author} {\bibfnamefont{J.~F.}\ \bibnamefont{Vibert}}, \bibinfo {author}
  {\bibfnamefont{M.}~\bibnamefont{Stiber}},\ and\ \bibinfo {author}
  {\bibfnamefont{S.}~\bibnamefont{Hanneton}},\ }%
  \bibfield{journal}{%
  \bibinfo {journal} {Neuroscience}\ }%
  \textbf{\bibinfo {volume} {68}},\ \bibinfo {pages} {657} (\bibinfo {year}
  {1995})%
  \bibAnnoteFile{NoStop}{SegundoVibertStiberHanneton1995a}%
\bibitem{SegundoVibertStiberHanneton1995b}%
  \BibitemOpen
  \bibfield{author}{%
  \bibinfo {author} {\bibfnamefont{J.~P.}\ \bibnamefont{Segundo}}, \bibinfo
  {author} {\bibfnamefont{M.}~\bibnamefont{Stiber}}, \bibinfo {author}
  {\bibfnamefont{J.~F.}\ \bibnamefont{Vibert}},\ and\ \bibinfo {author}
  {\bibfnamefont{S.}~\bibnamefont{Hanneton}},\ }%
  \bibfield{journal}{%
  \bibinfo {journal} {Neuroscience}\ }%
  \textbf{\bibinfo {volume} {68}},\ \bibinfo {pages} {693} (\bibinfo {year}
  {1995})%
  \bibAnnoteFile{NoStop}{SegundoVibertStiberHanneton1995b}%
\bibitem{SegundoVibertStiber1998}%
  \BibitemOpen
  \bibfield{author}{%
  \bibinfo {author} {\bibfnamefont{J.~P.}\ \bibnamefont{Segundo}}, \bibinfo
  {author} {\bibfnamefont{J.~F.}\ \bibnamefont{Vibert}},\ and\ \bibinfo
  {author} {\bibfnamefont{M.}~\bibnamefont{Stiber}},\ }%
  \bibfield{journal}{%
  \bibinfo {journal} {Neuroscience}\ }%
  \textbf{\bibinfo {volume} {87}},\ \bibinfo {pages} {15} (\bibinfo {year}
  {1998})%
  \bibAnnoteFile{NoStop}{SegundoVibertStiber1998}%
\bibitem{Dorval2005}%
  \BibitemOpen
  \bibfield{author}{%
  \bibinfo {author} {\bibfnamefont{A.~D.}\ \bibnamefont{Dorval},
  \bibfnamefont{Jr}}\ and\ \bibinfo {author} {\bibfnamefont{J.~A.}\
  \bibnamefont{White}},\ }%
  \bibfield{journal}{%
  \bibinfo {journal} {J. Neurosci.}\ }%
  \textbf{\bibinfo {volume} {25}},\ \bibinfo {pages} {10025} (\bibinfo {year}
  {2005})%
  \bibAnnoteFile{NoStop}{Dorval2005}%
\bibitem{Perkel1964}%
  \BibitemOpen
  \bibfield{author}{%
  \bibinfo {author} {\bibfnamefont{D.~H.}\ \bibnamefont{Perkel}}, \bibinfo
  {author} {\bibfnamefont{J.~H.}\ \bibnamefont{Schulman}}, \bibinfo {author}
  {\bibfnamefont{T.~H.}\ \bibnamefont{Bullock}}, \bibinfo {author}
  {\bibfnamefont{G.~P.}\ \bibnamefont{Moore}},\ and\ \bibinfo {author}
  {\bibfnamefont{J.~P.}\ \bibnamefont{Segundo}},\ }%
  \bibfield{journal}{%
  \bibinfo {journal} {Science}\ }%
  \textbf{\bibinfo {volume} {145}},\ \bibinfo {pages} {61} (\bibinfo {year}
  {1964})%
  \bibAnnoteFile{NoStop}{Perkel1964}%
\end{thebibliography}%

\clearpage

\begin{figure}[htbp]
\begin{center}
\includegraphics[scale=1]{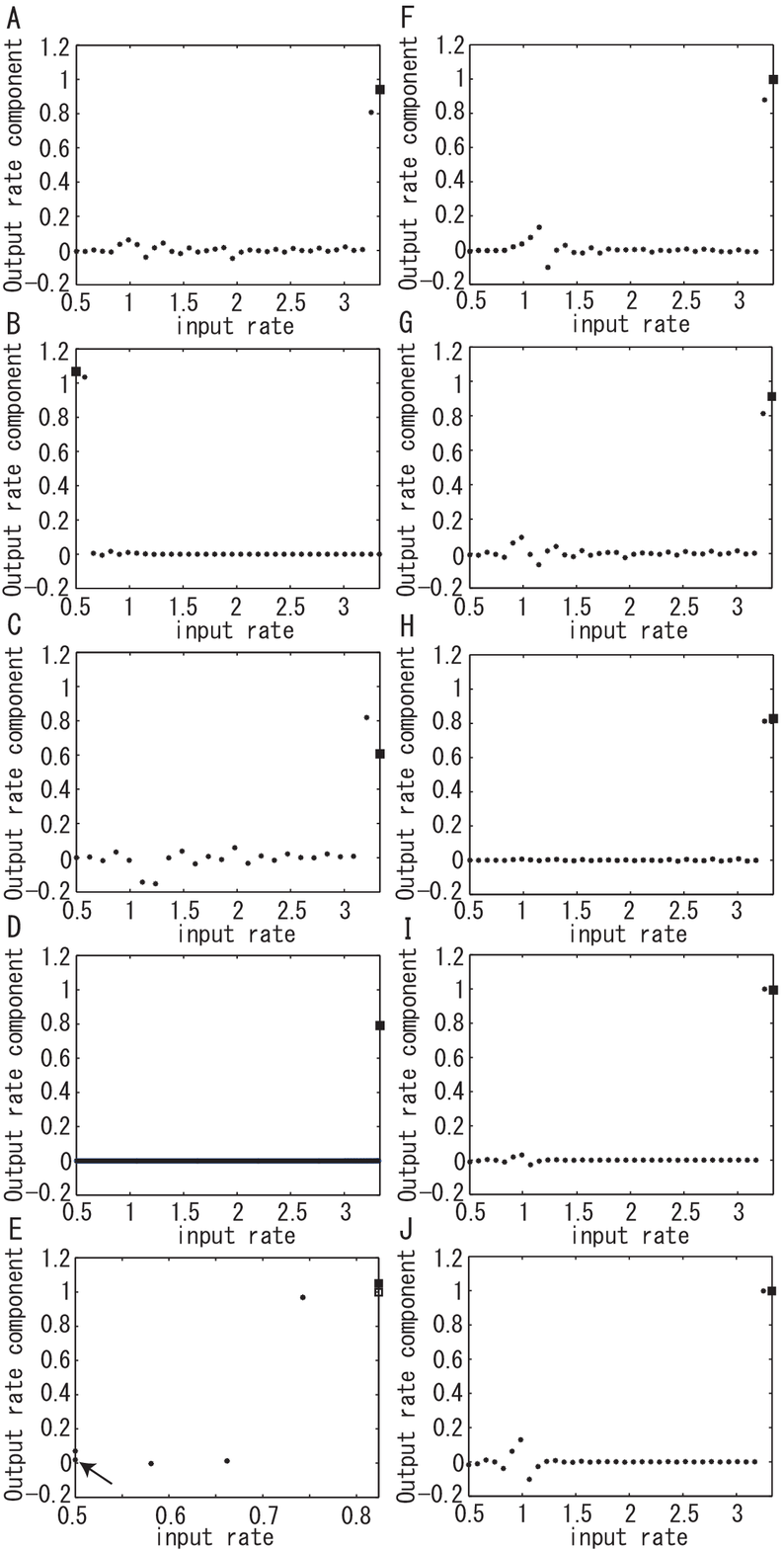}
\end{center}
\end{figure}

\begin{figure}[htbp]
\caption{ {\bf Dependence of the current instantaneous stochastic
rotation number on the past activity of the stochastic Poincar\'{e}
oscillator.} Each component of the current instantaneous stochastic
rotation number explained in Eq.~(\ref{eq:StoI4I3I2I1}) is plotted
as a function of the input rates. The filled squares (filled square and square in (E)) 
show the current instantaneous rotation number in each
panel. (A) Standard plot for starting and final input rates of
$f_{start} = 0.5$ and $f_{end}=1/0.3$, $N=35$ impulses with an
amplitude of $A=0.95$, a relaxation rate of $K=1$, and a noise
strength of $\epsilon=0.3$. The initial density $h_1(r_1,\phi_1)$
was a uniform distribution with a support of $(r_1,\phi_1) \in
(0,3.5]\times[0,1)$. In (B)--(J), the effects of various parameters
are calculated by varying one parameter while the other parameters
and initial density are the same as those in (A). The output
components as a function of the input rate for (B) $f_{start} =
1/0.3$ and $f_{end} = 0.5$ to investigate the effect of $f_{step}$
in Eq.~(\ref{eq:TimeVaryingImpulsesStep}), (C) $N=23$, (D) $N=500$. 
(There is a overlap the filled square and the asterisk),
and (E) $f_{start}=0.5$, $f_{end}=0.82381$, and $N=4$. In (E)
results for two initial densities are shown: the density in (A)
(asterisks) and a uniform distribution with a support of
$(r_1,\phi_1) \in (0,3.5]\times[0.25,0.75)$ (asterisks indicated by an arrow). 
There are three overlaps between two responses, and the filled square and 
square show the corresponding current instantaneous rates. Plots
are also shown for (F) $K=\infty$, (G) $K=0.6$, (H) $\epsilon = 0.6$, (I)
$A=0.25$ and (J) $A=0.45$.} \label{Figure_label8}
\end{figure}

\begin{figure}[htbp]
\begin{center}
\includegraphics[scale=1]{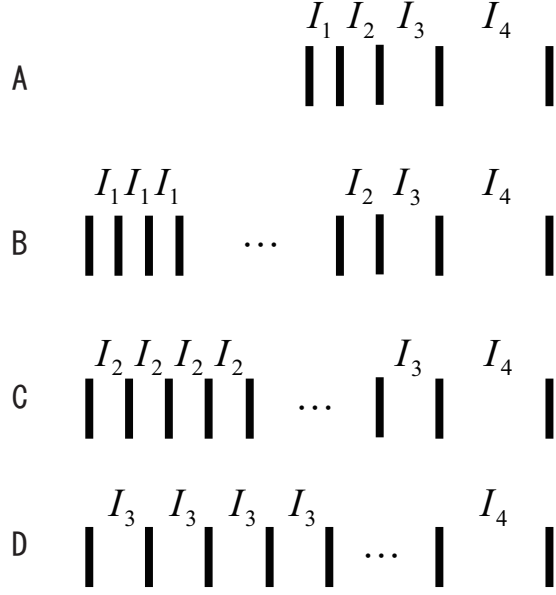}
\end{center}
\caption{ {\bf Experimental procedure to determine the components of
the current instantaneous stochastic rotation number.} (A) Five
impulses determined by a constant impulse amplitude $A$ and four
interimpulse intervals of $I_1$, $I_2$, $I_3$, and $I_4$ are added
repeatedly to a spontaneously firing neuron with a
recovery period. We decompose the current firing rate, i.e., the
firing rate during $I_4$, into the components determined by the past
activity of the neuron.  Impulses to measure (B) the summation of
the first, second, and third terms, (C) the summation of the first
and second terms, and (D) the first term in
Eq.~(\ref{eq:StoI4I3I2I1}). The firing rate during $I_4$ needs to be
measured by repeatedly adding the impulses to the neuron with a
recovery period. Subtraction using the measured firing rates during
$I_4$ in (A)--(D) leads to the decomposition of the firing rate
during $I_4$ in (A) into the components in
Eq.~(\ref{eq:StoI4I3I2I1}). See the discussion for a detailed
explanation. } \label{Figure_label9}
\end{figure}

\end{document}